\documentclass[a4paper,11pt]{article}
\pdfoutput=1 

\usepackage{jheppub} 

\usepackage[T1]{fontenc} 

\usepackage{comment}

\newcommand{\abs}[1]{\left |#1 \right|}

\newcommand{\tr}{\operatorname{tr}}
\newcommand{\diag}{\operatorname{diag}}

\newcommand{\adj}{\operatorname{adj}}
\newcommand{\MP}{M_{\rm P}}

\title{\huge Pseudo-Goldstone dark matter in a radiative inverse seesaw scenario}

\author[1]{K. Kannike,\note{Corresponding author.}}
\author[]{A. Kubarski,}
\author[]{L. Marzola}
\author[]{and A. Racioppi}

\affiliation[]{National Institute of Chemical Physics and Biophysics, R\"avala 10, 10143, Tallinn, Estonia}

\emailAdd{kristjan.kannike@cern.ch}
\emailAdd{aleksei.kubarski@ut.ee}
\emailAdd{luca.marzola@cern.ch}
\emailAdd{antonio.racioppi@kbfi.ee}

\abstract{We consider a scale-invariant inverse seesaw model with dynamical breaking of gauge symmetry and lepton number. In some regions of the parameter space, the Majoron -- the pseudo-Goldstone of lepton number breaking -- is a viable dark matter candidate. The bound on the Majoron decay rate implies a very large dilaton vacuum expectation value, which also results in a suppression of other dark matter couplings. Because of that, the observed dark matter relic abundance can only be matched via the freeze-in mechanism. The scalar field which gives mass to heavy neutrinos can play the role of the inflaton, resulting in a tensor-to-scalar ratio $r \lesssim 0.01$ for metric inflation and $r \lesssim 0.21$ for Palatini gravity.}

\begin{document}
\maketitle
\flushbottom

\section{Introduction}
\label{sec:intro}

The discovery of the Higgs boson~\cite{ATLAS:2012yve, CMS:2012qbp} is the last of a series of triumphs that the Standard Model (SM) of particle physics achieved during almost 50 years of dedicated experimental testing. In spite of its successes, the SM remains an incomplete theory of Nature as several questions motivated either by experiments or by theoretical arguments remain unanswered. In particular, the lightness of the neutrino mass scale, the cosmological inflation and the nature of dark matter (DM) seem to call for new particle physics phenomena beyond the boundaries of the SM.

When comparing the masses of known fermions, neutrinos stand out for their lightness. Within the SM, once right-handed (RH) neutrinos are considered, the lightness of the neutrino mass scale can be matched upon a tuning of the corresponding Yukawa couplings. Although from the experimental point of view this is a valid solution to the problem of neutrino masses, the resulting hierarchy in the Higgs boson couplings -- spanning more than 10 orders of magnitude -- does generally not sit well with the aesthetics of theoreticians. In order to reduce the graveness of this `small hierarchy problem', more involved mechanisms of neutrino mass generation were consequently proposed. In the type I seesaw mechanism~\cite{Gell-Mann:1979vob, Minkowski:1977sc, Mohapatra:1979ia, Schechter:1980gr, Yanagida:1979as}, for instance, the suppression of the light neutrino masses results from a large Majorana mass term of RH neutrinos, taken to transform trivially under the SM gauge group. In the type II scenario~\cite{Schechter:1980gr, Mohapatra:1980yp, Wetterich:1981bx, Magg:1980ut, Cheng:1980qt}, instead, a new scalar in the triplet representation of $SU(2)_L$ is used to generate the neutrino scale in place of RH neutrinos.  The type III seesaw scenario~\cite{Foot:1988aq}, finally, generalises the type I case by letting RH neutrinos transform as a triplet of $SU(2)_L$. Regardless of the specifics of the construction, a common trait of seesaw models is that the typical scale of new physics is the large scale causing the suppression of neutrino masses, hence the new particles involved are typically inaccessible to current collider experiments. Because of this, in the present paper we choose to focus on the inverse seesaw mechanism, which, inversely, is a mechanism of \textit{low-scale} seesaw~\cite{Boucenna:2014zba}. According to this framework, the neutrino masses are generated by a small Majorana mass scale $\mu$ that breaks lepton number. Since in the limit $\mu \rightarrow 0$ lepton number is conserved and the symmetry of the theory is enhanced, the Majorana scale is naturally small \cite{tHooft:1979rat}.

A further unappealing feature of the SM manifests itself again as a hierarchy problem, this time concerning the ratio between the Higgs boson mass and Planck scale. According to common lore, quantum corrections to the Higgs boson two-point function should receive large contributions from heavy particles between the electroweak scale and Planck scale, driving this ratio towards unity in absence of fine-tuning. In relation to this, given that the Higgs boson mass  is the only dimensionful parameter of the SM~\cite{Bardeen:1995kv}, we propose to use the Coleman-Weinberg (CW) mechanism~\cite{Coleman:1973jx} to generate this parameter in a classically scale-invariant model, thereby partially alleviating the severity of this hierarchy problem. In the case where several scalar fields are involved, it is customary to use the Gildener-Weinberg method~\cite{Gildener:1976ih} to find the approximate direction in field space along which the CW mechanism operates.

We thus discuss a scale-invariant realization of the SM with an inverse seesaw mechanism, in line with earlier attempts presented in Refs.~\cite{Hempfling:1996ht, Meissner:2006zh, Meissner:2009gs, Meissner:2007xv, Foot:2007as, Foot:2007ay}; later studies include~\cite{Hambye:2007vf, Holthausen:2009uc, Iso:2009nw, Iso:2009ss, Alexander-Nunneley:2010tyr, Hur:2011sv, Gabrielli:2013hma, Farzinnia:2013pga, Holthausen:2013ota, Davoudiasl:2014pya, Benic:2014aga, Altmannshofer:2014vra, Farzinnia:2014xia, Lindner:2014oea, Kubo:2014ova, Karam:2015jta, Humbert:2015epa, Humbert:2015yva, Karam:2016rsz, Helmboldt:2016mpi, Ahriche:2016cio, Ahriche:2016ixu, Brdar:2018vjq,Kierkla:2022odc}. Several possibilities for classically scale-invariant ways to generate neutrino masses are discussed in~\citep{Lindner:2014oea}, including inverse seesaw models in which the $B - L$ symmetry is gauged, e.g.~\cite{Humbert:2015yva,Rojas:2017sih,Biswas:2018yus,Nomura:2018mwr,Abada:2021yot,Mohapatra:2023aei}. In contrast, we do not gauge the $B - L$ symmetry, so the scale invariance is solely broken by scalar interactions.

The outline of our scenario is as follows: lepton number is broken spontaneously by the vacuum expectation value (VEV) of a singlet scalar with $L = 2$~\citep{Gonzalez-Garcia:1988okv}, giving rise to a Majorana mass term for new fermion fields as well as to a massless Goldstone boson -- the Majoron~\citep{Chikashige:1980ui, Schechter:1981cv,Mandal:2021acg}. On top of this, we consider also a small \textit{explicit} breaking of lepton number, which renders our Majoron a massive \textit{pseudo}-Goldstone boson. This allows us to use the Majoron as a DM candidate (see e.g.~\cite{Rojas:2017sih,Brune:2018sab,Biggio:2023gtm}) with the related phenomenology constraining the parameter space of the model. In regard of this, we point out that the properties of our Majoron field differ from those supported by the simplest case of pseudo-Goldstone DM with a $U(1)$-breaking mass term~\citep{Gross:2017dan}, in particular, by the presence of further contributions entering the determination of the Majoron mass.

Interestingly, the framework we adopt allows us to draw connections also with another open problem in particle physics, related to the mechanism of cosmological inflation. The observations of the cosmic microwave background show that the universe is flat and homogenous at large scales, to a point that an explanation seems needed. The most natural and simplest one relies on an early stage of exponential expansion of the universe, the cosmic inflation~\cite{Starobinsky:1980te,Guth:1980zm,Linde:1981mu,Albrecht:1982wi}, which has also the merits to produce the inhomogeneities that seeded the growth of large scale structure. Such an exponential expansion can be driven by a scalar field, commonly dubbed the inflaton, with a potential that dominates the energy budget (see~\cite{Martin:2013tda} for a comprehensive review). Curiously, potentials \emph{\`a la} Coleman-Weinberg have been often used in this context (e.g. \cite{Kannike:2014mia,Kannike:2015apa,Kannike:2015kda,Kannike:2016jfs,Marzola:2016xgb,Artymowski:2016dlz,Racioppi:2017spw,Karam:2017rpw,Kannike:2018zwn,Karam:2018mft,Racioppi:2018zoy,Racioppi:2019jsp,Gialamas:2020snr,Racioppi:2021ynx,Karam:2023haj} and Refs. therein), with them being even among the first proposed inflationary scenarios~\cite{Linde:1981mu, Albrecht:1982wi}.

The purpose of this paper is then that of cross-correlating the phenomenologies of neutrino physics, dark matter and inflation within the boundaries of a scale invariant construction. We determine the parameter space in which radiative symmetry breaking, neutrino mass and Majoron DM relic density simultaneously satisfy the corresponding experimental bounds. To perform the required parameter space scan, we adopt a Markov Chain Monte Carlo (MCMC) method taking into account various theoretical and experimental constraints, such as perturbativity, the invisible decay width of the Higgs boson, and the experimental constraints on the DM relic density and neutrino sector. For the study of radiative symmetry breaking, we employ the nice matrix formalism given in ref.~\cite{Kannike:2019upf} and further developed in ref.~\cite{Kannike:2020ppf}. 

The paper is structured as follows. The model we adopt in our investigation is described in section~\ref{sec:model}, whereas the radiative symmetry breaking dynamics is investigated in section~\ref{sec:sym:br}. Details of neutrino mass generation are given in section~\ref{sec:nu:mass}. Section \ref{sec:results} presents the details of the scan, with theoretical and experimental constraints given in section~\ref{sec:constraints}, results for scalar and neutrino sector in section~\ref{sec:parameter:space}, lepton flavour violation in section~\ref{sec:lfv}, DM in section~\ref{sec:dm} and the inflation phenomenology in section~\ref{sec:inflation}. We conclude in section~\ref{sec:conclusions}. Appendix~\ref{sec:vacuumstab} contains the bounded-from-below conditions for the scalar quartic couplings, appendix~\ref{sec:betafunct} the $\beta$-functions of gauge, Yukawa and scalar couplings,  appendix~\ref{sec:flat:direction} the matrix Gildener-Weinberg formalism, and appendix~\ref{appendix:inflation} on the details of inflationary parameters.

\section{The Model}
\label{sec:model}

\begin{table}[tb]
\caption{Relevant field content of the model.}
\begin{center}
  \begin{tabular}{ | l | c | c | c | }
    \hline
    Particle 	& $SU(2)$ 	& $Y_W$	& $L$ \\ \hline
    $\ell$ 		& \textbf{2} 			& $-1$ 	& $+1$ \\ \hline
    $H$ 		& \textbf{2} 			& $+1$ 	& $0$ \\ \hline
    $N_R$ 		& \textbf{1} 			& $0$ 	& $+1$\\ \hline
    $S_L$ 		& \textbf{1} 			& $0$ 	& $+1$\\ \hline
    $\sigma$ 	& \textbf{1} 			& $0$ 	& $-2$\\ \hline
    $\rho$ 		& \textbf{1} 			& $0$ 	& $0$ \\
    \hline
  \end{tabular} 
\end{center}
\label{tab:fields}
\end{table}%

We introduce new scalar fields in order to generate the Higgs mass term and neutrino masses dynamically. The scalar sector of the model then comprises a complex singlet $\sigma$ and a real singlet $\rho$, in addition to the SM Higgs doublet $H$. Implementing the inverse seesaw also requires the inclusion, besides SM left-handed leptons $\ell$, of two new two-component singlet fermion fields, $N_R$ and $S_L$, with lepton number $L(N_R) = L(S_L) = +1$. The relevant field content of the model is summarised in table~\ref{tab:fields}. The part of the Yukawa term Lagrangian that regulates the physics of neutrinos is

    \begin{equation}
	\mathcal{L}_{\nu} = 
	-Y_D^{ij} \bar{\ell}_L^j i \tau_2 H^* N_R^i
	- Y_{NS}^{ij} \rho \bar{N}_{R_i} S_{L_j}
	- Y_{S}^{ij} \sigma \tilde{S}_{L_i} S_{L_j}
	+ \text{H.c.},
	\end{equation}
	where $\tilde{S_L}$ is defined as $\tilde{S_L}=S_L^T C^{-1}$, where $C$ is the charge conjugation matrix. The scale-invariant scalar potential is given by 
	\begin{equation}
		\begin{split}
			V &= 
			\lambda_{H} \abs{H}^{4} 
			+ \lambda_\sigma \abs{\sigma}^{4} 
			+ \frac{1}{4} \lambda_\rho \rho^4 
			+ \lambda_{H\sigma} \abs{H}^{2} \abs{\sigma}^{2}
			+ \frac{1}{2} \lambda_{H\rho} \abs{H}^{2} \rho^2 
			+ \frac{1}{2} \lambda_{\rho\sigma} \abs{\sigma}^{2} \rho^2 \\
			&+ \frac{1}{2} \lambda'_{H\sigma} ( \sigma^2 + \sigma^{*^2} ) \abs{H}^{2}
			+ \frac{1}{4} \lambda'_{\rho\sigma} ( \sigma^2 + \sigma^{*^2} )\rho^2
			+ \frac{1}{2} \lambda'_{\sigma} (\sigma^4 + \sigma^{*^4} )
			\\
			&
			+ \frac{1}{2} \lambda''_{\sigma} ( \sigma^2 + \sigma^{*^2} ) \abs{\sigma}^{2},
		\end{split}
\label{eq:V:0}
\end{equation}
which is invariant under the discrete symmetry transformations $\rho \to -\rho$, $\sigma \to -\sigma$ and $\sigma \to \sigma^{*}$. The last of these parities is broken by the Yukawa sector, leading eventually to the decay of the Majoron. The primed couplings, instead, break lepton number explicitly and only in the limit $\lambda_{H\sigma}' = \lambda_{\rho\sigma}' = \lambda_\sigma' = \lambda_\sigma'' = 0$ the Majoron is massless.\footnote{For simplicity, we assume that only the scalar potential contains terms that explicitly break lepton number.} We thus expect these couplings to be naturally small in comparison to other interactions.

We parametrise the scalar fields as
\begin{equation}
  \abs{H}^{2} = \frac{h^{2}}{2}, \qquad \sigma = \frac{R + i J}{\sqrt{2}},
\end{equation}
where the pseudo-Goldstone boson $J$ is the Majoron. Writing the potential \eqref{eq:V:0} in terms of these field components, we have
\begin{equation}
\begin{split}
  V &= \frac{1}{4} \lambda_{H} h^{4} + \frac{1}{4} \lambda_{\rho} \rho^{4}
   + \frac{1}{4} \lambda_{R} R^{4} + \frac{1}{4} \lambda_{J} J^{4} + \frac{1}{4}  \lambda_{RJ} R^{2} J^{2}
    + \frac{1}{4} \lambda_{H\rho} h^{2} \rho^{2}
    \\
   & + \frac{1}{4} \lambda_{HR} h^{2} R^{2}
   + \frac{1}{4} \lambda_{HJ} h^{2} J^{2} 
   + \frac{1}{4}  \lambda_{\rho R} R^{2} \rho^{2}
   + \frac{1}{4}  \lambda_{\rho J} J^{2} \rho^{2},
\end{split}
\label{eq:V:rp}
\end{equation}
where
\begin{align}
  \lambda_{R} &= \lambda_{\sigma} + \lambda'_{\sigma} + \lambda''_{\sigma},
  \\
  \lambda_{J} &= \lambda_{\sigma} + \lambda'_{\sigma} - \lambda''_{\sigma},
  \\
  \lambda_{RJ} &= 2 (\lambda_{\sigma} - 3 \lambda'_{\sigma}),
  \\
  \lambda_{HR} &= \lambda_{H\sigma} + \lambda'_{H\sigma},
  \\
  \lambda_{HJ} &= \lambda_{H\sigma} - \lambda'_{H\sigma},
  \\
  \lambda_{\rho R} &= \lambda_{\rho\sigma} + \lambda'_{\rho\sigma},
  \\
  \lambda_{\rho J} &= \lambda_{\rho\sigma} - \lambda'_{\rho\sigma}.
\end{align}
We henceforth refer to this reparametrization and, in order to ensure that the lepton-number breaking quartic couplings be small, we impose $\abs{\lambda'_{H\sigma}} \ll \abs{\lambda_{H\sigma}}$, $\abs{\lambda'_{\rho\sigma}} \ll \abs{\lambda_{\rho\sigma}}$ and $\abs{\lambda'_{\sigma}}, \abs{\lambda''_{\sigma}} \ll \abs{\lambda_{\sigma}}$. For the reparametrised couplings this implies
\begin{align}
  \abs{\lambda_{HR} - \lambda_{HJ}} &\ll \abs{\lambda_{HR} + \lambda_{HJ}},
  \label{eq:smalllep1}
  \\
    \abs{\lambda_{\rho R} - \lambda_{\rho J}} &\ll \abs{\lambda_{\rho R} + \lambda_{\rho J}},
\label{eq:smalllep2}
  \\
  \abs{\lambda_{R} + \lambda_{J} - \lambda_{RJ}} &\ll \abs{3 (\lambda_{R} + \lambda_{J}) + \lambda_{RJ}}, \label{eq:smalllep3}
  \\
  4 \abs{\lambda_{R} - \lambda_{J}} &\ll \abs{3 (\lambda_{R} + \lambda_{J}) + \lambda_{RJ}}.
  \label{eq:smalllep4}
\end{align}

The scalar spectrum of the model contains three CP-even mass eigenstates: the SM-like Higgs boson $h_1$ with mass $m_1 = 125.1$~GeV, a heavy Higgs boson $h_2$ with a TeV-scale or higher mass and the dilaton $h_3$, usually the lightest of the three. The mass eigenstate $h_1$ is closely aligned with the gauge eigenstate $h$, while the other scalars exhibit larger mixing. The single physical CP-odd scalar, the Majoron $J$, typically has sub-GeV mass to avoid too fast DM decay. The Higgs boson $h$ obtains the VEV $v_h = 246.22$~GeV. The scalar $R$ typically has the largest VEV, while $\rho$ has a VEV between the two. The mass eigenstates in the neutrino sector include the three light neutrinos and six heavy neutrinos with masses $M_i$ of $\mathcal{O}(\mathrm{TeV})$. The scalar mass spectrum is detailed in Section~\ref{sec:sym:br} and the neutrino mass spectrum in Section~\ref{sec:nu:mass}.

\section{Radiative symmetry breaking and scalar masses}
\label{sec:sym:br}

In order to calculate the flat direction and relate the particle masses to quartic couplings, we use the convenient matrix formalism given in ref.~\cite{Kannike:2019upf} and further developed in ref.~\cite{Kannike:2020ppf}.

The tree-level potential~\eqref{eq:V:rp} can be written as
\begin{equation}
  V = \left(\mathbf{\Phi}^{\circ 2}\right)^T \mathbf{\Lambda} \mathbf{\Phi}^{\circ 2},
\label{eq:V}
\end{equation} 
where the field vector $\mathbf{\Phi} = (h,\rho,R,J)^{T}$ and the coupling matrix is given by
\begin{equation}
  \mathbf{\Lambda} = \frac{1}{4}
  \begin{pmatrix}
    \lambda_{H} & 
    \frac{1}{2} \lambda_{H\rho} & 
    \frac{1}{2} \lambda_{HR} & 
    \frac{1}{2} \lambda_{HJ} 
    \\
    \frac{1}{2} \lambda_{H\rho} & 
    \lambda_{\rho} & 
    \frac{1}{2} \lambda_{\rho R} & 
    \frac{1}{2} \lambda_{\rho J}
    \\
    \frac{1}{2} \lambda_{HR} &
    \frac{1}{2} \lambda_{\rho R} &
    \lambda_{R} &
   \frac{1}{2} \lambda_{RJ}
    \\
   \frac{1}{2} \lambda_{HJ}   &
    \frac{1}{2} \lambda_{\rho J} &
    \frac{1}{2} \lambda_{RJ} &
    \lambda_{J}
  \end{pmatrix}.
\label{eq:lam:mtrx}
\end{equation}
To write the potential \eqref{eq:V}, we use the Hadamard product, $\circ$, defined as the element-wise product of two tensors of equal rank, for instance: $(\mathbf{A} \circ \mathbf{B})_{ij} = A_{ij} B_{ij}$ for $\mathbf{A}$ and $\textbf{B}$ matrices. Similarly, the Hadamard power of a matrix is defined as $(\mathbf{A}^{\circ n})_{ij} = A^n_{ij}$ and so the Hadamard square of the field vector $\mathbf{\Phi}$ simply is $(\mathbf{\Phi}^{\circ 2})_{i} = (\mathbf{\Phi}_{i})^{2}$.

As shown in ref.~\cite{Kannike:2019upf}, the Hadamard square of the unit vector that tracks the flat direction of the scalar potential in field space is given by
\begin{equation}
  \mathbf{n}^{\circ 2} = \frac{\adj(\mathbf{\Lambda}) \, \mathbf{e}}{\mathbf{e}^T \! \adj(\mathbf{\Lambda}) \, \mathbf{e}},
\label{eq:flat:direction:sol}
\end{equation} 
where $\adj(\mathbf{\Lambda})$ is the adjugate of the coupling matrix, satisfying $\adj(\mathbf{\Lambda}) \mathbf{\Lambda} = \det (\mathbf{\Lambda}) \mathbf{I}$, and $\mathbf{e} = (1, \ldots, 1)^T$ is a vector of ones with the same length as $\mathbf{\Phi}$. A detailed derivation of Eq.~\eqref{eq:flat:direction:sol} is given in appendix~\ref{sec:flat:direction}.

We can write the tree-level scalar mass matrix $\mathbf{m}^{2}_{S}$ and the coupling matrix $\mathbf{\Lambda}$ in the block form  
\begin{equation}
  \mathbf{m}^{2}_{S} = 
  \begin{pmatrix}
    \mathbf{m}^{2}_{HH} & \mathbf{m}^{2}_{HA}
    \\
   (\mathbf{m}^{2}_{HA})^{T} & \mathbf{m}^{2}_{AA}
  \end{pmatrix},
  \quad
  \mathbf{\Lambda} = 
  \begin{pmatrix}
    \mathbf{\Lambda}_{HH} & \mathbf{\Lambda}_{HA}
    \\
    (\mathbf{\Lambda}_{HA})^{T} &\mathbf{\Lambda}_{AA}
  \end{pmatrix},
\label{eq:lambda:block}
\end{equation}
where the index $H$ denotes CP-even and the index $A$ CP-odd states. Since the Majoron $J$ does not develop a VEV and there is no mixing between the CP-even and CP-odd states, the mass matrix must be block-diagonal. The flat direction also partitions as $\mathbf{n} = (\mathbf{n}_H, 0)^T$, where $\mathbf{n}_H$ is the non-zero CP-even part. We then have a $3 \times 3$ block $\mathbf{m}^{2}_{HH}$ of CP-even fields and a remaining $1 \times 1$ block $\mathbf{m}^{2}_{AA}$ containing the Majoron mass. Necessarily $\mathbf{m}^{2}_{HA} = \mathbf{0}$. By solving for part of the quartic couplings in terms of the masses and mixing angles, we obtain
\begin{equation}
  \mathbf{m}^{2}_{S} = v_{\varphi}^{2}  
  \begin{pmatrix}
    8 \mathbf{\Lambda}_{HH} \circ (\mathbf{n}_{H} \mathbf{n}_{H}^{T}) & \mathbf{0}
    \\
    \mathbf{0}^{T} &4 \mathbf{\Lambda}_{HA}^{T} \mathbf{n}_{H}^{\circ 2}
  \end{pmatrix},
\label{eq:hessian:part:zero}
\end{equation}
where $v_{\varphi}$ is the dilaton VEV.
Therefore, the $3 \times 3$ block $\mathbf{\Lambda}_{HH}$ of the quartic coupling matrix is related to the tree-level mass matrix as
\begin{equation}
  \mathbf{\Lambda}_{HH} = \frac{1}{8 v_{\varphi}^{2}} \mathbf{m}_{HH}^{2} \circ (\mathbf{n}_{H} \mathbf{n}_{H}^{T})^{\circ -1}.
\label{eq:lambda:from:hessian}
\end{equation}
The flat-direction condition $\det(\mathbf{\Lambda}_{HH}) = 0$ is imposed by setting the tree-level mass of the dilaton -- the pseudo-Goldstone of classical scale invariance -- to zero.


We define the mixing matrix $\mathbf{O}$ diagonalizing the CP-even scalar masses as $\mathbf{O}^{T} \mathbf{m}_{HH}^{2} \mathbf{O} = \diag (m_{1}^{2}, m_{2}^{2}, m_{3}^{2})$, where $m_{i}^{2}$ are the tree-level mass eigenvalues corresponding to the CP-even eigenstates $h_{i}$ and the mixing matrix $\mathbf{O}$ is given by
\begin{equation}
\begin{split}
  \mathbf{O} &= \mathbf{O}_{23} \mathbf{O}_{31} \mathbf{O}_{12} = 
  \\
  &=
  \begin{pmatrix}
  c_{12} c_{13} & s_{12} c_{13}  & s_{13}
  \\ 
  -s_{12} c_{23} - c_{12} s_{23} s_{13} & c_{12} c_{23} - s_{12} s_{23} s_{13} & s_{23} c_{13} 
  \\
  s_{12} s_{23} - c_{12} c_{23} s_{13}  & -c_{12} s_{23} - s_{12} c_{23} s_{13}  & c_{23} c_{13} 
  \end{pmatrix},
\end{split}
\end{equation}
where $\mathbf{O}_{ij}(\alpha)$ is the rotation matrix in the $ij$ subspace by the angle $\alpha_{ij}$. For the sake of brevity we denoted $s_{ij} \equiv \operatorname{sin}\alpha_{ij}$, $c_{ij} \equiv \cos \alpha_{ij}$. The gauge and mass eigenstates are then related as
\begin{equation}
  \begin{pmatrix}
    h
    \\
    \rho
    \\
    R
  \end{pmatrix}
  =
  \mathbf{O} 
    \begin{pmatrix}
    h_{1}
    \\
    h_{2}
    \\
    h_{3}
  \end{pmatrix},
\end{equation}
where, in our conventions, the ordering of states is fixed regardless of the mass hierarchy. We point out that the Higgs boson cannot be identified with the dilaton. In fact, having a dilaton mass match the measured Higgs mass requires large $m_2$ or $m_3$ and, since all the other VEVs are generally smaller than the Higgs VEV, this would result in non-perturbative values for some scalar couplings.

We thus opt to identify $h_{1}$ with the SM Higgs boson, setting $m_{1} = 125.1$~GeV, and the dilaton $\varphi$ with the field $h_{3}$, which has a vanishing tree-level mass $m_{3}$. Then the flat direction is given by the last column of the mixing matrix $\mathbf{O}$, that is
\begin{equation}
  \mathbf{n}_{H} = 
  \begin{pmatrix}
  s_{13} \\ s_{23} c_{13}  \\ c_{23} c_{13} 
  \end{pmatrix}.
\end{equation}
By definition, the CP-even VEVs are given by $(v_h, v_\rho, v_R)^T = v_\varphi \mathbf{n}_H.$
 
The coupling matrices $\mathbf{\Lambda}_{HA}$ and $\mathbf{\Lambda}_{AA}$ are under-determined, except that we must require that the Majoron mass be positive,
\begin{align}
  m_{J}^{2} &= 4 \mathbf{\Lambda}_{HA}^{T} \mathbf{n}_{H}^{\circ 2} \, v_{\varphi}^{2} \\
  			&= \frac{1}{2} \left( \lambda_{HJ} v_h^2 + \lambda_{\rho J} v_\rho^2 + \lambda_{RJ} v_R^2 \right)
  > 0.
\end{align}
On top of that, we set the Majoron self-coupling $\mathbf{\Lambda}_{AA} = \lambda_{J} > 0$ so that the potential is bounded from below.

In the minimum direction, the dilaton potential is given by
\begin{equation}
  V(\varphi) = \frac{\mathbb{B}}{v_{\varphi}^{4}} \varphi^{4} \left( \ln \frac{\varphi^{2}}{v_{\varphi}^{2}} - \frac{1}{2} \right),
  \label{eq:Vvarphi}
\end{equation}
where
\begin{equation}
  64 \pi^{2} \mathbb{B} = 6 m_{W}^{4} + 3 m_{Z}^{4} + \sum_{i = 1}^{3} m_{i}^{4} + m_{J}^{4} - 12 m_{t}^{4} - 2 \sum_{i = 1}^{3} M_{i}^{4},
  \label{eq:Bcoef}
\end{equation}
and the dilaton mass then is
\begin{equation}
  M_{\varphi}^{2} = 8 \frac{\mathbb{B}}{v_{\varphi}^{2}}.
  \label{eq:dmass}
\end{equation}
Note that the requirement $M_{\varphi}^{2} > 0$ yields an upper bound on the new heavy fermion states $M_i$, once the Majoron mass $m_J$ is specified.

At high temperature, thermal corrections~\cite{Marzola:2017jzl} to the potential in eq.~\eqref{eq:Vvarphi} become relevant. The potential is then given by
\begin{equation}
  V(\varphi, T) = V(\varphi) + C T^2 \varphi^2,
\end{equation}
where
\begin{equation}
  C = \frac{1}{12 v_\varphi^2} \left( 6 m_{W}^{2} + 3 m_{Z}^{2} + \sum_{i = 1}^{3} m_{i}^{2} + m_{J}^{2} + 6 m_{t}^{2} + \sum_{i = 1}^{3} M_{i}^{2}\right).
\end{equation}
The dominant mass is the scalar mass $m_2$, hence thermal corrections are relevant for $T > m_2$. Although, as we will see, the reheating temperature $T_R$ typically satisfies this bound, for the sake of the DM relic density calculation we can neglect  temperatures larger than $m_2$ because the VEV of the thermal potential essentially vanishes and  the DM abundance produced in this temperature range consequently dilutes as radiation.

\section{Neutrino sector}
\label{sec:nu:mass}

We adopt the inverse seesaw mechanism to explain the lightness of the neutrino mass scale and, to that end, introduce new heavy neutral leptons $N_R$ and $S_L$. In the basis $\nu_{L}$, $N_{R}^c$, $S_{L}$, the full $9 \times 9$ neutrino mass matrix is given by the block matrix
\begin{equation}
  \mathbf{M}_{\nu} = 
  \begin{pmatrix}
    \mathbf{0} & \mathbf{m}_{D}^{T} & \mathbf{0}
    \\
    \mathbf{m}_{D} & \mathbf{0} & \mathbf{M}^{T}
    \\
    \mathbf{0} & \mathbf{M} & \boldsymbol{\mu}
  \end{pmatrix},
  \label{eq:Mnu}
\end{equation}
where $\mathbf{m}_{D} = \mathbf{Y}_{D} v_{h}$, $\mathbf{M} = \mathbf{Y}_{NS} v_{\rho}$ and $\boldsymbol{\mu} = \mathbf{Y}_{S} v_{R}$. Note that $\mathbf{m}_D$ and $\mathbf{M}$ are arbitrary $3 \times 3$ Dirac mass matrices and $\boldsymbol{\mu}$ is a $3 \times 3$ Majorana matrix. We can assume, without loss of generality, that the matrix $\mathbf{M}$ is diagonal. Because the $\boldsymbol{\mu}$ term breaks lepton number spontaneously, we expect it to be naturally small~\citep{Gonzalez-Garcia:1988okv}. In the limit $\boldsymbol{\mu} \rightarrow 0$, the light neutrinos are massless since the $U(1)$ symmetry associated to total lepton number conservation is not broken. There is, however, lepton flavour violation~\citep{Gonzalez-Garcia:1991brm, Bernabeu:1987gr}, together with CP-violation if the couplings are complex~\citep{Rius:1989gk, Branco:1989bn}. We will describe the constraints from lepton flavour violation in section~\ref{sec:lfv}.

In the limit where $\boldsymbol{\mu}, \mathbf{m}_{D} \ll \mathbf{M}$, we have that the light neutrino mass matrix is approximately given by \cite{Gonzalez-Garcia:1988okv}
\begin{equation}
  \mathbf{m}_{\nu} = \mathbf{m}_{D} \frac{1}{\mathbf{M}} \boldsymbol{\mu} \frac{1}{\mathbf{M}^T}  \mathbf{m}_{D}^{T}.
  \label{eq:mnu}
\end{equation} 

We define the mixing matrix $\mathbf{W}$ that approximately diagonalizes the neutrino mass matrix \eqref{eq:Mnu}, $\mathbf{W}^T \mathbf{M}_\nu \mathbf{W} \approx \diag(N_i)$, as 

\begin{equation}
\label{eq:Wmatrix}
\mathbf{W}=
\begin{pmatrix}	
 	\mathbf{U} -  \frac{1}{2} \frac{1}{\mathbf{M}} \mathbf{m}_D  \mathbf{m}_D^{ \dagger} \frac{1}{\mathbf{M}}\mathbf{U}
	& \frac{1}{\sqrt{2}} \mathbf{m}_D^{^\dagger}\frac{1}{\mathbf{M}}
	&  \frac{i}{\sqrt{2}} \mathbf{m}_D^{^\dagger}\frac{1}{\mathbf{M}} \\
	0
	& \frac{1}{\sqrt{2}} \mathbf{I} 
	&  -\frac{i}{\sqrt{2}} \mathbf{I}   \\
	- \frac{1}{\mathbf{M}}\mathbf{m}_D \mathbf{U}
	& \frac{1}{\sqrt{2}} \mathbf{I}  - \frac{1}{2\sqrt{2}}\frac{1}{\mathbf{M}} \mathbf{m}_D  \mathbf{m}_D^{ \dagger} \frac{1}{\mathbf{M}}
	&  \frac{i}{\sqrt{2}} \mathbf{I}  - \frac{i}{2\sqrt{2}}\frac{1}{\mathbf{M}} \mathbf{m}_D  \mathbf{m}_D^{ \dagger} \frac{1}{\mathbf{M}} \\
\end{pmatrix}.
\end{equation}

The Casas-Ibarra parametrisation~\cite{Casas:2001sr} of the neutrino Yukawa couplings obtained from Eq.~\eqref{eq:mnu} is \cite{Deppisch:2004fa}
\begin{equation}
\mathbf{Y}_D = \mathbf{M}^T \mathbf{U}_\mu \frac{1}{\sqrt{\mathbf{D}_\mu}} \mathbf{R} \sqrt{\mathbf{D}_\kappa} \mathbf{U}^\dagger,
\label{eq:ciparam}
\end{equation}
where $\mathbf{U}_{\mu}$ is the mixing matrix of the Majorana submatrix $\boldsymbol{\mu}$, the matrix $\mathbf{D}_{\mu}$ is the diagonalised $\boldsymbol{\mu}$ matrix, the matrix $\mathbf{D}_{\kappa}$ is the diagonalised $\boldsymbol{\kappa} = \mathbf{m}_{\nu}/v_{h}$, $\mathbf{U}$ is the matrix that diagonalises $\mathbf{m}_{\nu}$, and $\mathbf{R}$ is an arbitrary $3 \times 3$ complex orthogonal matrix.

The seesaw expansion parameter $\epsilon \sim \| \mathbf{m}_{D}/\mathbf{M} \|$ characterises the strength of unitarity violation and can be on the order of $10^{-2}$~\citep{Bazzocchi:2009kc, Forero:2011pc}. The deviation from unitarity is typically
of $\mathcal{O}(\epsilon^2)$.

The eigenvalues, mixing angles and phases of the light neutrino mass matrix $\mathbf{m}_{\nu}$ must satisfy the experimental bounds on the neutrino mass differences and absolute scale. Current experiments still allow for two orderings of neutrino mass eigenstates: the normal ordering (NO) for which $m_{\nu 1} < m_{\nu 2} < m_{\nu 3}$ and the inverted ordering (IO), corresponding to $m_{\nu 3} < m_{\nu 1} < m_{\nu 2}$. In the case of NO, the best fit of experimental data yields $\sin^2 \theta_{12} = 0.304$, $\sin^2 \theta_{13} = 0.022$ and $\sin^2 \theta_{23} = 0.573$, $\delta_{\rm CP}=197^\circ$, $\Delta m_{21}^2=7.42 \cdot 10^{-5} \rm ~eV^2$, $\Delta m_{31}^2=2.517 \cdot 10^{-3} \rm ~eV^2$~\citep{Esteban:2020cvm} (For IO, the parameters that differ from the NO case are
 $\sin^2 \theta_{23} = 0.575$, $\delta_{\rm CP}=282^\circ$ and $\Delta m_{32}^2=-2.498 \cdot 10^{-3} \rm eV^2$.) The sum of the light neutrino masses, or equivalently the neutrino absolute scale, must satisfy an upper bound due to cosmological observations: $\sum m_{\nu i} < 0.12$~eV \cite{RoyChoudhury:2018gay}. As the light neutrinos described by the inverse seesaw mechanism are Majorana particles, the light neutrino mass matrix $\mathbf{m}_{\nu}$ possesses also two additional Majorana phases which are not constrained by experiments. 

\section{Results}
\label{sec:results}

We perform two scans: 1) in a general MCMC scan, we find the parameter space which generates neutrino masses and satisfies all the constraints but does not try to explain other open questions, 2) in a second grid scan we address, besides neutrino masses, also cosmic inflation and DM with the Majoron as our DM candidate. The general parameter space is described in subsections \ref{sec:constraints}, \ref{sec:parameter:space} and \ref{sec:lfv}, the dark matter and inflation phenomenology in subsections \ref{sec:dm} and \ref{sec:inflation}.

\subsection{Constraints and MCMC scanning strategy}
\label{sec:constraints}

We now briefly review the constraints that we use in our exploration of the parameter space of the model. First of all, in order for the model to be perturbative, we require the scalar couplings to be less than $\pi$ in absolute value at the EW scale. We then track their running, specified in the $\beta$-functions of appendix~\ref{sec:betafunct}, to assess perturbativity up to the Planck scale $M_{\rm P} = 1.22 \times 10^{19}$~GeV. The initial values we use in our RGE running calculations are given in Ref.~\cite{Buttazzo:2013uya}.

In the limit of large field values the scalar potential must be bounded from below, in order to have a stable finite potential minimum. Because the potential~\eqref{eq:V:0} is biquadratic in the norms of fields  -- and the potential~\eqref{eq:V:rp} in real field components, -- the vacuum stability conditions can be enforced by requiring copositivity of the coupling matrix \eqref{eq:lam:mtrx}~\citep{Kannike:2012pe}. For the improved effective potential we use the renormalisation group equations (RGE) given in appendix~\ref{sec:betafunct} to check up to which scale the conditions given in appendix~\ref{sec:vacuumstab} are satisfied.\footnote{Notice that in order to have a radiative minimum, the conditions must be violated in a \textit{finite} range of fields below the flat direction scale. Above the latter, they must be respected.}

Measurements of the Yukawa and gauge couplings of the Higgs boson, together with direct searches for new light degrees of freedom at the LHC, put constraints on the mixing of this particle with other scalars. Because in our case the SM-like Higgs boson is identified with $h_1$ upon diagonalization of the CP-even scalar sector, the current measurements result in a lower bound on $c^2_{12} c^2_{13}$~\citep{Robens:2016xkb}. Similarly, the Higgs boson signal analyses constrain the possible (invisible) decays of this state into new particles. In the present framework, these processes are active if the Majoron mass, the mass of $h_2$, or the dilaton $h_3$ mass are below $m_h/2$. For instance, the partial width for the Higgs boson decay into Majorons is given by
\begin{equation}
  \Gamma_{h_{1} \to JJ} = \frac{g_{1JJ}^{2}}{32 \pi m_{h}} \sqrt{1 - \frac{4 m_{J}^{2}}{m_{h}^{2}}},
  \label{eq:Higgs:invis}
\end{equation}
where $g_{1JJ}$ is the coefficient of the $h_1 JJ$ term in the potential.
 The invisible decay rate $\Gamma_{h_{1} \to JJ}$, entering the invisible branching ratio $\mathrm{BR}_{\rm inv} = \Gamma_{h_{1} \to JJ}/(\Gamma_{h_{1} \to JJ} + \Gamma_{h_{1} \to \rm{SM}})$, must satisfy $\mathrm{BR}_{\rm inv} < 0.11$ (ATLAS) \cite{ATLAS:2023tkt} or $\mathrm{BR}_{\rm inv} < 0.15$ (CMS) \cite{CMS:2023sdw}, of which we use the stricter ATLAS bound.

Finally, we check that the seesaw expansion parameter $\epsilon$ that characterises unitarity violation be smaller than $10^{-2}$ \citep{Bazzocchi:2009kc, Forero:2011pc}.

The mentioned constraints enter the Markov Chain Monte Carlo (MCMC) scan that we use to explore the parameter space of the model, comprising 28 free parameters. Our scanning strategy consists of the following four steps: 

\begin{itemize}

\item First, we generate scalar mixing angles and the $h_2$ mass. No limitations are set on the ranges of these parameters, apart from the bound $c_{12} c_{13}>0.89$ due to the measured Higgs signals~\citep{Robens:2016xkb}. The quartic couplings of CP-even fields are calculated from Eq.~\eqref{eq:lambda:from:hessian} and points which violate perturbativity are discarded.

\item Secondly, the coupling constants associated with the Majoron $J$ are generated. To ensure that the conditions 
\eqref{eq:smalllep1}, \eqref{eq:smalllep2}, \eqref{eq:smalllep3} and \eqref{eq:smalllep4} yielding lepton number breaking are satisfied, these couplings have to be of size similar to the analogous $R$ couplings. (We required that the left-hand-side of each inequality is at least an order of magnitude smaller than the right-hand-side.) Therefore, a more natural choice for the MCMC scan is to consider as parameters the ratios $\lambda_{HJ}/\lambda_{HR}$, $\lambda_{\rho J}/\lambda_{\rho R}$ and $\lambda_{RJ}/\lambda_{R}$. The Majoron self-coupling $\lambda_J$, entering only the lepton number breaking, can be set by the means of Eqs.~\eqref{eq:smalllep3} and \eqref{eq:smalllep4} for given values of $\lambda_{RJ}$ and $\lambda_{R}$. Any set of couplings that violates pertubativity is discarded.

\item In the third step, we generate the eigenvalues of $\mathbf{M}$. We discard any point that yields an unphysical dilaton mass ($M_{\varphi}^2 < 0$) and also check whether the parameters satisfy the bounds on scalar mixing angles $c_{12}$ and $c_{13}$, as interpolated from data in Ref.~\citep{Robens:2016xkb}. 

\item In the last step, we generate all the remaining free parameters: the eigenvalues of $\boldsymbol{\mu}$, mixing angles and phases of $\boldsymbol{\mu}$, Majorana phases of light neutrinos, mass of lightest neutrino, real and imaginary parts of angles in $\mathbf{R}$. We consider only the normal ordering of neutrino masses and vary the lightest neutrino mass $m_{\nu 1}$ in a logarithmic distribution between $10^{-6}$~eV and $0.03$~eV. The other masses are set by using the best-fit values of $\Delta m_{ij}^2$ given at the end of Section~\ref{sec:nu:mass}; the neutrino mixing angles are also set to the corresponding best fit values. We check that the light neutrino masses satisfy the cosmological bound $\sum m_{\nu i}<0.12$~eV. The Majorana phases vary in the range $[0,2\pi)$. To ensure the presence of a small explicit lepton number breaking, the eigenvalues of the $\boldsymbol{\mu}$ matrix are taken in the $1-10$~eV range. All points violating perturbativity of Yukawa interactions (couplings larger than $\sqrt{\pi}$) or leading to an hierarchy different from $\max [( \mathbf{m}_D)_{ij}] < \min ( \mathbf{M}_{ij}/10 )$, are discarded.
\end{itemize}

Because we do not attempt to infer the probability distributions of the involved parameters, but only the shape of the allowed parameter space, the target function was set to maximise points in the boundary region. Namely, the MCMC was set up to favour points with higher values of couplings as shown in table~\ref{tab:scansteps}. In the first and second steps, we used a linear distribution of absolute maximum values for the relevant scalar couplings. In the third step, we use instead a uniform distribution across the boundary region. In the last step, the target function is a linear distribution of absolute maximum value for all the new Yukawa couplings. We did not normalize the distributions as only relative values between points were needed.

\begin{table}[tb]
\caption{Scanned parameter space of each step with target distribution if scanned point does not violate any of the restrictions. $\lambda^{\rm CP-even}$ refers to couplings between CP-even states.}
\begin{center}
  \begin{tabular}{ | c | c | c |}
    \hline
    Step & Scanned parameters & Target distribution 	\\
    \hline
    1 & $\sin \alpha_{12}$, $\sin \alpha_{13}$, $\sin \alpha_{23}$, $m_2$  & $\text{Max}(|\lambda_{H}|,|\lambda_{H\rho}|, |\lambda_{HR}|, |\lambda_{\rho}|, |\lambda_{\rho R}|, |\lambda_{R}|)$ \\
    \hline
    2 & $\lambda_{HJ}$, $\lambda_{\rho J}$, $\lambda_{RJ}$  & $\text{Max}(|\lambda_{HJ}|, |\lambda_{\rho J}|,|\lambda_{RJ}|)$ \\
    \hline
    3 & eigenvalues of $\mathbf{M}$  & Uniform distribution \\
    \hline
    4 & $\boldsymbol{\mu}$, $\mathbf{R}$, $m_{\nu 1}$, light neutrino Majorana phases & $\text{Max}(|\text{elements of } \mathbf{Y}_{D,S,NS}|)$ \\
    \hline
  \end{tabular} 
\end{center}
\label{tab:scansteps}
\end{table}%

\subsection{Scalar and neutrino sector}
\label{sec:parameter:space}

\begin{figure}[tbp]
\centering
\includegraphics[scale=0.8]{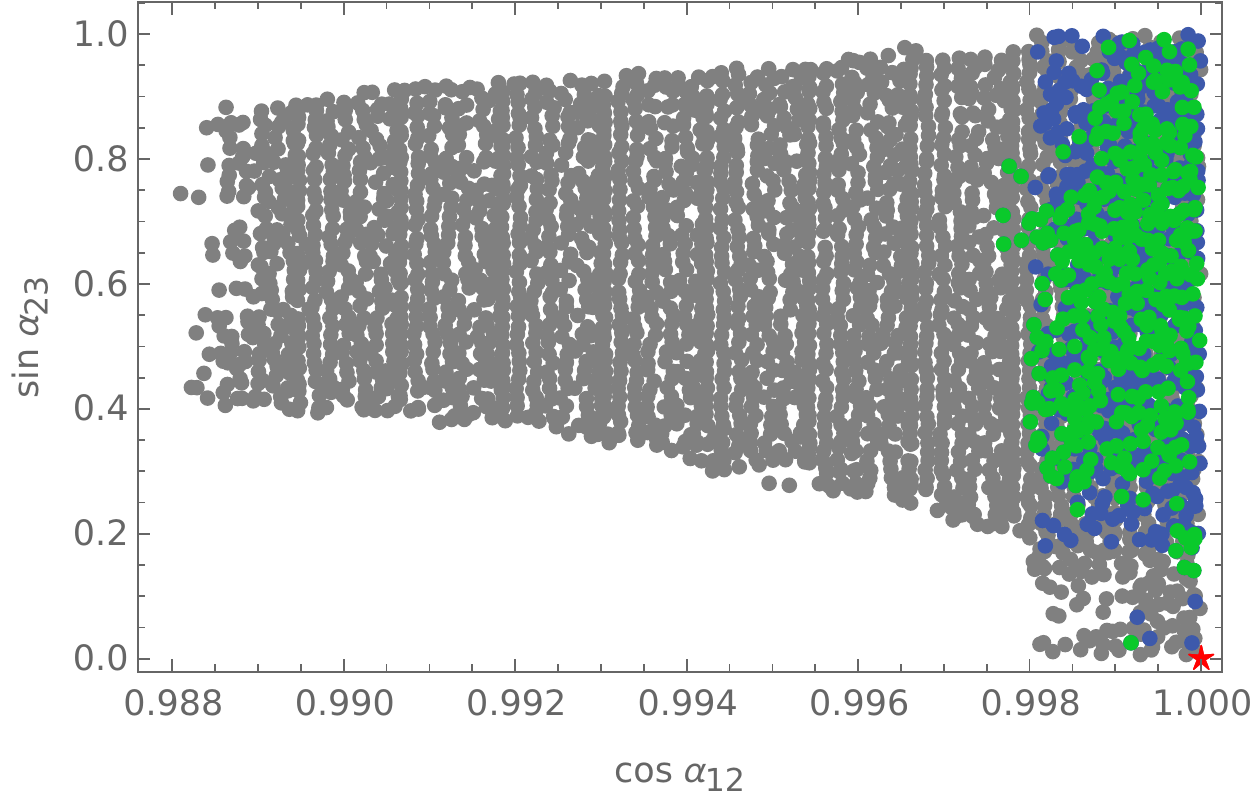}
\caption{The allowed parameter space in the $\sin \alpha_{23}$ vs. $\cos \alpha_{12}$ plane of scalar mixing angles. Generic points are shown in grey, points perturbative up to the Planck scale in blue, and points that satisfy vacuum stability conditions at the Planck scale in green. The red star marks the parameter space where both DM relic density and cosmic inflation can be explained.}
\label{fig:c12s23}
\end{figure}

\begin{figure}[ptb]
\begin{center}
  \includegraphics[scale=0.55]{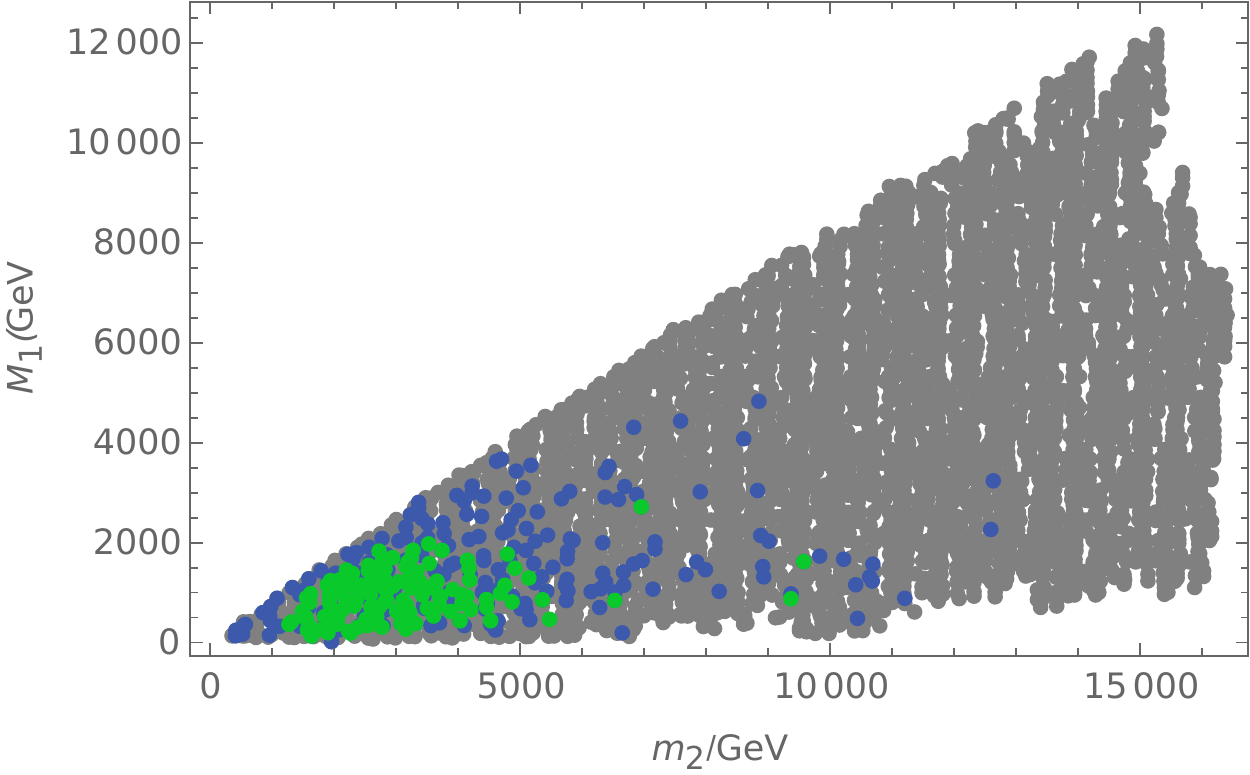}\qquad
  \includegraphics[scale=0.55]{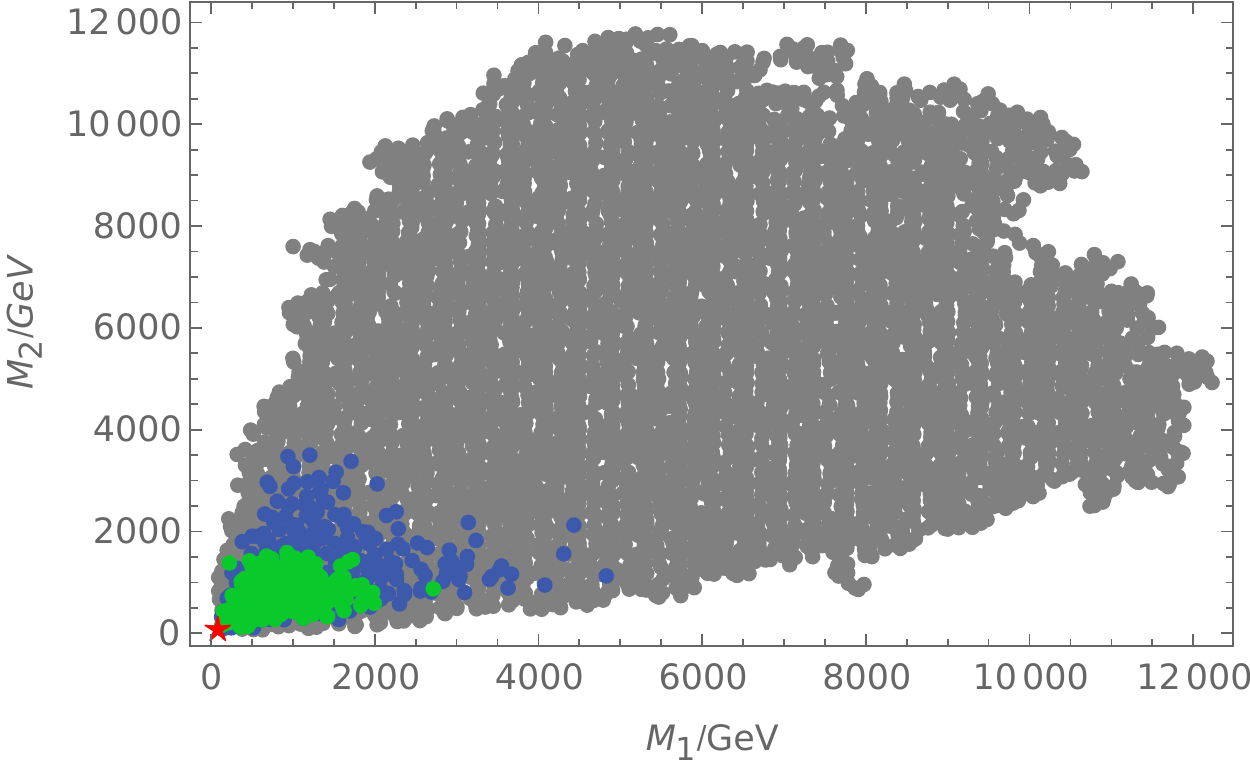}
  \\
  \vspace{2em}
   \includegraphics[scale=0.55]{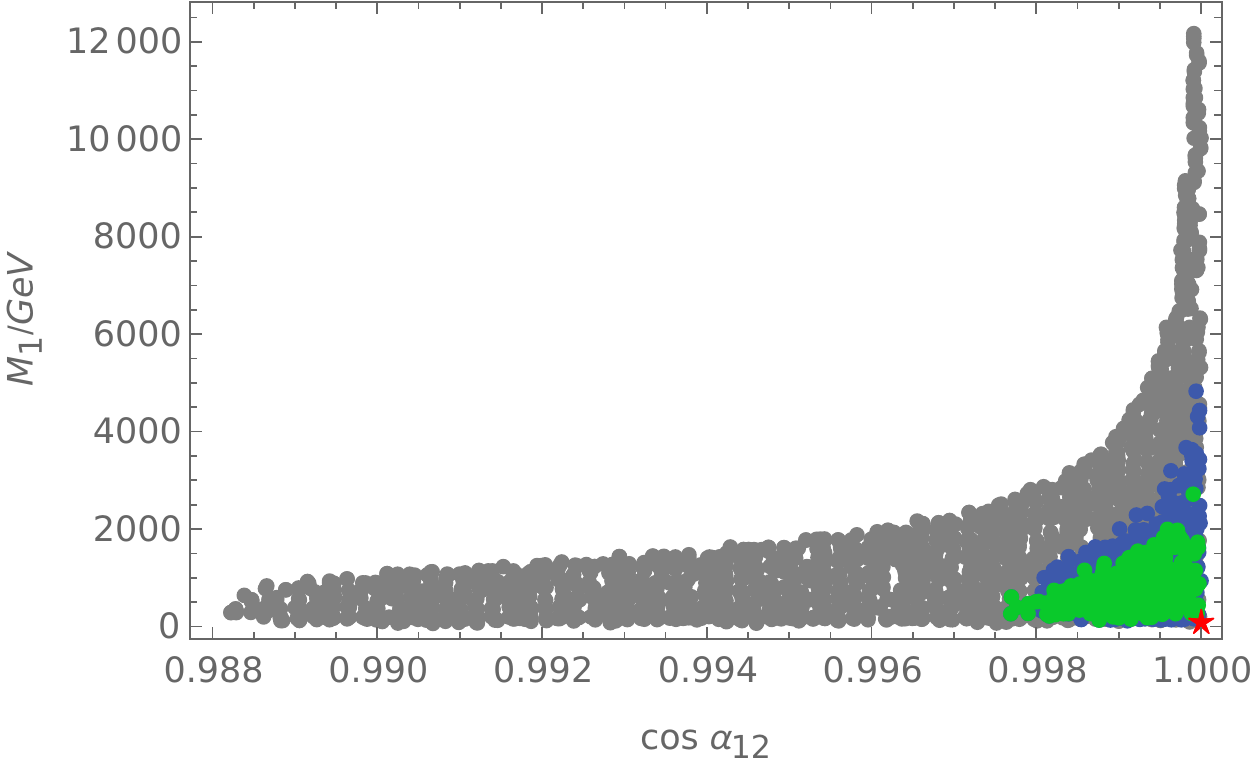}\qquad
  \includegraphics[scale=0.55]{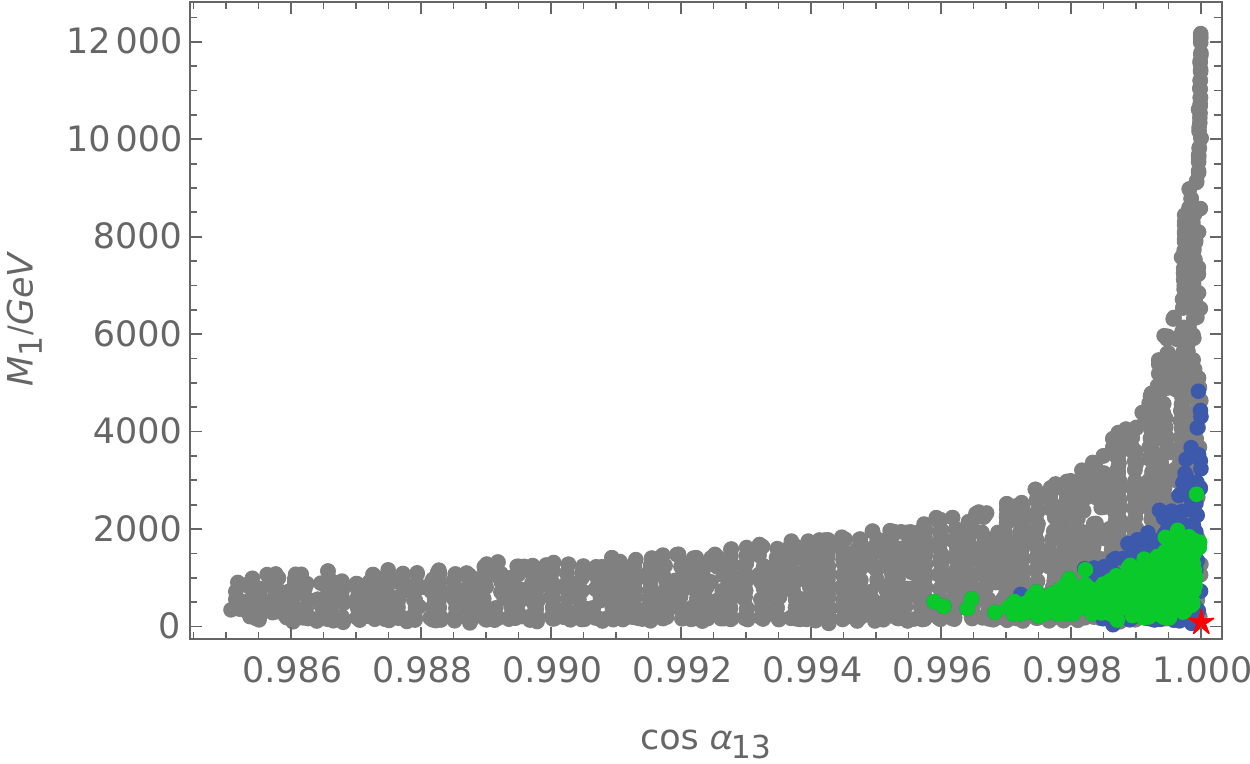}
\caption{Allowed parameter space of the heavy neutrino mass eigenvalue $M_1$ relative to $m_2$, $M_2$, $\cos  \alpha_{12}$ and $\cos \alpha_{13}$. Generic points are shown in grey, points perturbative up to the Planck scale in blue, and points that satisfy vacuum stability conditions at the Planck scale in green. The red star marks the parameter space where both DM relic density and cosmic inflation can be explained.}
\label{fig:M1vsoth}
\end{center}
\end{figure}

The results of the scan show that the free parameters in the fermion sector have negligible correlation with each other and are mostly left unconstrained. In particular, there are enough free parameters to achieve either the NO or IO of the neutrino masses by simply choosing suitable values of the angles in the $\mathbf{R}$ and $\boldsymbol{\mu}$ matrices that enter the Casas-Ibarra parametrisation~\eqref{eq:ciparam}. Within the framework, NO and IO can then be obtained from each other upon a reordering of the eigenvalues of $\mathbf{M}$ or $\boldsymbol{\mu}$. For this reason, we choose not to perform a separate scan for the IO case. The allowed excursion of the lightest neutrino mass does not affect the remaining parameter space as only the upper bounds to $\mathbf{Y}_D$ are constrained, which are reached first by the heaviest generation.

An example of allowed range for the mixing angles can be seen in Fig.~\ref{fig:c12s23}, where generic points in the allowed parameter space are shown in grey, the points that are perturbative up to the Planck scale are blue and the points that satisfy the vacuum stability conditions are green. The red star marks the parameter space where both the DM relic density is in agreement with data and cosmic inflation can be explained by a single scalar that is not the Higgs boson. We see that very large $\alpha_{23}$ angles are disfavoured due to non-perturbativity of the coupling constants $\lambda_{R}$ and $\lambda_{HR}$. Although the condition $\mathbf{m}_D \ll \mathbf{M}$ favours larger $\alpha_{23}$ angles, it limits only very small angles where $\lambda_{\rho}$ and $\lambda_{H\rho}$ are also not perturbative. The cut-off close to $\cos  \alpha_{12} = 0.989$ is due to the constraint on the decay width of the Higgs boson into dilatons. Smaller values of the $\alpha_{12}$ and $\alpha_{13}$ angles increase $v_\rho$ and $v_\sigma$ and, therefore, reduce the values of coupling constants and allow larger $m_2$ values.

The Higgs invisible width constraint, applied to the $h \to JJ$ process~\eqref{eq:Higgs:invis}, does not noticeably influence the shape of the allowed parameter space. In particular, in the case where the Majoron is a dark matter candidate, we find that the corresponding branching ratio is negligible.

The heavy neutrino mass matrix $\mathbf{M}$ is linked to the scalar sector mostly through the $\mathbb{B}$ coefficient described in Eq.~\eqref{eq:Bcoef}. Note that eigenvalues of $\mathbf{M}$ contribute to $\mathbb{B}$ with negative sign, whereas $\mathbb{B}$ must be positive to ensure that the dilaton mass \eqref{eq:dmass} is physical. Therefore, given the scalar masses $m_2$ and $m_J$ (which give a positive contribution to $\mathbb{B}$), the size of the eigenvalues of $\mathbf{M}$ is limited by $\mathbb{B} > 0$ with $\mathbb{B}$ given in Eq.~\eqref{eq:Bcoef}. In the main part of the scan, where we do not consider Majoron DM, its mass can exceed 7~TeV. It is still, however, subject to the conditions \eqref{eq:smalllep1}-\eqref{eq:smalllep4} and remains always below the value of $m_2$ in the performed scan. This implies that the upper limits of the eigenvalues of $\mathbf{M}$ are directly limited by $m_2$, as it can be seen in the first panel of Fig.~\ref{fig:M1vsoth}. To allow for a large value of $m_2$, the mixing angles $\alpha_{12}$ and $\alpha_{13}$ must be small. In the opposite case, in fact, the VEVs of the new scalar fields would be too small to generate large scalar masses maintaining the perturbativity of the involved couplings. Since the eigenvalues of $\mathbf{M}$ are bounded from above by $M_\varphi^2 > 0$ for given scalar masses, the same condition is necessary to have large values of $M_i$, as it can seen in the third and fourth panels of Fig.~\ref{fig:M1vsoth}. Also, as it can be seen in second panel, the hierarchy in the eigenvalues of $\mathbf{M}$ cannot be substantial due to the requirement that these quantities be all at least one order of magnitude larger than the eigenvalues of $\mathbf{m}_D$.

To satisfy the hierarchy $\mathbf{M} \gg \mathbf{m}_{D}$ in the neutrino mass matrix, it is more desirable to have a large mixing angle $\alpha_{23}$. For values close to $\sin \alpha_{23}=0.64$, however, the $R$ self-coupling $\lambda_R$ becomes non-perturbative in correspondence of a large scalar mass $m_2$.

\subsection{Lepton number violation}
\label{sec:lfv}

In order to calculate the rates of lepton flavour violating processes, 
we use the results of Ref.~\cite{Forero:2011pc}. The branching ratio for the process $l_i \rightarrow l_j \gamma$ is given by
\begin{equation}
	Br(l_i \rightarrow l_j \gamma) = \frac{\alpha_W^2 s_W^2 }{256\pi^2} \frac{m_{l_i}^5}{M^4_W} \frac{1}{\Gamma_{l_i}} \abs{G_{ij}^W}^2,
\end{equation}
where the loop functions are
\begin{equation}
	G^W_{ij} =\sum_{k=1}^9 K_{ik}^* K_{jk} G_\gamma^W \left( \frac{m_{N_k}^2}{M^2_W} \right),
\end{equation}
and
\begin{equation}
	G^W_\gamma(x) = \frac{1}{12(1-x)^4} \left( 10 - 43x + 78x^2 -49x^3 +18x^3 \ln x + 4x^4 \right).
\end{equation}

The effective lepton mixing matrix $\mathbf{K}$ is \cite{Schechter:1980gr}
\begin{equation}
	K_{ba} = \sum_{c=1}^3 \Omega_{cb} W_{ca} = (\mathbf{U} - \frac{1}{2} \frac{1}{\mathbf{M}} \mathbf{m}_D \mathbf{m}_D^\dagger \frac{1}{\mathbf{M}} \mathbf{U}, \frac{1}{\sqrt{2}} \mathbf{m}_D^\dagger \frac{1}{\mathbf{M}}, \frac{i}{\sqrt{2}} \mathbf{m}_D^\dagger \frac{1}{\mathbf{M}}),
\end{equation}
where we take the charged lepton mixing matrix $\mathbf{\Omega}$ to be unit matrix, and the mixing matrix $\mathbf{W}$ is given by \eqref{eq:Wmatrix}.
 The strictest limit on flavour violating decay branching ratio comes from MEG collaboration~\cite{MEG:2016leq}:
\begin{equation}
	Br(\mu \rightarrow e \gamma) < 4.2\cdot 10^{-13}.
\end{equation}
We also checked that constraints coming from the Belle experiment~\cite{Belle:2021ysv} $Br(\tau \rightarrow \mu \gamma) < 4.2\cdot 10^{-8}$ and $Br(\tau \rightarrow e \gamma) < 5.6\cdot 10^{-8}$ are also satisfied, finding that these bound do not significantly further restrict  the parameter space.

\subsection{Dark matter}
\label{sec:dm}

\begin{table}[tb]
\caption{Benchmark points for Majoron DM freeze-in. The first three points are from an unrestricted scan which showed no strong constrains on parameter space from DM relic density and last five points have $\lambda_{\rho}>10^{-12}$ which is necessary for $\rho$ to be inflaton candidate. For the inflationary parameters, the upper index $M$ stands for metric and $P$ for Palatini.}
\begin{center}
  \begin{tabular}{ | c | c | c | c | c | c | c | c | c |}
    \hline
    $\sin \alpha_{12} $ & $0.01$ & $0.03$ & $0.17$ & $10^{-4}$ & $10^{-5}$ & $10^{-5}$ & $10^{-5}$ & $5\cdot 10^{-6}$	\\
    \hline
    $\sin \alpha_{13}$ & $10^{-11}$  & $10^{-11}$ & $10^{-12}$ & $10^{-11}$ & $10^{-11}$  & $10^{-12}$ & $10^{-11}$ & $10^{-12}$\\
    \hline
    $\sin \alpha_{23}$ & $0.31$ & $0.71$ & $0.99$ & $0.005$ & $0.01$ & $0.005$ & $10^{-4}$ & $0.005$\\
    \hline
    $m_2/\rm GeV$ & $5000$ & $5000$ & $1000$ & $7\cdot 10^5$ & $3\cdot 10^6$ & $7\cdot 10^{6}$ & $9\cdot 10^{6}$ & $10^{7}$\\
    \hline
    $m_J/\rm GeV$ & $0.6$ & $0.01$ & $0.5$ & $0.1$ & $0.01$ & $0.1$ & $1$ & $0.1$\\
    \hline
    $M_{\varphi}/\rm GeV$ & $10^{-7}$ & $10^{-7}$ & $5\cdot 10^{-10}$ & $0.002$ & $0.04$ & $0.02$ & $0.37$ & $0.05$\\
    \hline
    $M_{1}/\rm GeV$ & $136$ & $145$ & $84$ & $23$ & $22$ & $22$ & $7$ & $47$\\
    \hline
    $M_{2}/\rm GeV$ & $135$ & $137$ & $81$ & $23$ & $23$ & $22$ & $6$ & $46$\\
    \hline
    $M_{3}/\rm GeV$ & $133$ & $153$ & $78$ & $26$ & $23$ & $22$ & $6$ & $43$\\
    \hline
    $T_{\rm R}/\rm GeV$ & -- & -- & -- & $7\cdot 10^9$ & $2\cdot 10^{10}$ & $10^{10}$ & $2\cdot 10^{10}$ & $4 \cdot 10^{10}$\\
    \hline
    $n_s^M$ & -- & -- & -- & $0.964$ & $0.964$ & $0.964$ & $0.967$ & $0.966$\\
    \hline
    $r^M$ & -- & -- & -- & $0.0075$ & $0.0075$ & $0.0039$ & $0.0050$ & $0.0042$\\
    \hline
    $N_e^M$ & -- & -- & -- & $53$ & $53$ & $53$ & $54$ & $54$\\
    \hline
    $n_s^P$ & -- & -- & -- & $0.970$ & $0.970$ & $0.971$ & $0.972$ & $0.977$\\
    \hline
    $r^P$ & -- & -- & -- & $0.018$ & $0.018$ & $0.0014$ & $0.0052$ & $6.1\cdot 10^{-4}$\\
    \hline
    $N_e^P$ & -- & -- & -- & $54$ & $54$ & $52$ & $53$ & $53$\\
    \hline
  \end{tabular} 
\end{center}
\label{tab:freezeout}
\end{table}%

As for dark matter phenomenology, the bound on the DM relic density $\Omega_{c}h^2 = 0.120 \pm 0.001$~\citep{Planck:2018vyg} must  be satisfied in the parameter space region where the Majoron acts as a dark matter candidate. In regard of this, the Majoron decay rate into light neutrinos,
\begin{equation}
  \Gamma_{J \to \nu_i \bar{\nu}_i} = \frac{m_{\nu i}^2 m_J}{4\pi v_R^2},
\label{eq:DM:Gamma}
\end{equation}
must be strongly suppressed so that the lifetime of DM can be much longer than the age of the universe~\cite{Arguelles:2022nbl}. Concretely, we require that the lifetime be longer than $10^{25}$~s.

DM can also decay into charged leptons through effective couplings induced at the loop level, with corresponding decay rates given by \cite{Garcia-Cely:2017oco,Heeck:2019guh}
\begin{equation}
\label{eq:DM:Gammaf}
\begin{split}
  \Gamma_{J \to q\bar{q}} &\simeq \frac{3}{8\pi} \abs{g_{Jqq}^P}^2 m_J,\\
  \Gamma_{J \to \ell \bar{\ell'}} &\simeq 
  \frac{1}{8\pi} \left( \abs{g_{J\ell \ell'}^P}^2 +  \abs{g_{J\ell \ell'}^S}^2 \right) m_J,
\end{split}
\end{equation}
where
\begin{equation}
\label{eq:leptong}
\begin{split}
g^P_{J\ell \ell'} &\simeq \frac{m_\ell + m_{\ell'}}{16\pi^2 v} \left( \delta_{\ell \ell'} T_3^\ell \tr \mathbf{K}' + \mathbf{K}'_{\ell \ell'} \right),
\\
g^S_{J\ell \ell'} &\simeq \frac{m_{\ell'} + m_\ell}{16\pi^2 v} \mathbf{K}'_{\ell \ell'},\\
g^P_{Jqq'} &\simeq \frac{m_q}{8\pi^2 v} \delta_{qq'} T_3^q \tr \mathbf{K}'.
\end{split}
\end{equation}
Here, $T_3^{d,\ell} = -T_3^u = -\frac{1}{2}$ and the matrix $\mathbf{K}'$, with elements $\mathbf{K}'_{\ell \ell'}$,  is defined as
\begin{equation}
\label{eq:matrixK}
	\mathbf{K}' =  \frac{\textbf{m}_D \textbf{m}_D^\dagger}{v_h v_R}.
\end{equation}
We require that the lifetime of DM to be at least order of magnitude longer than age of the universe.

To calculate the DM relic density and the spin-independent direct detection cross section, we prepared the model with the {\tt FeynRules 2} package~\citep{Alloul:2013bka} for export into the {\tt CalcHEP} format and used the {\tt micrOMEGAs} 5.3 code~\citep{Belanger:2020gnr,Alguero:2022inz}.

The constraint on the DM decay rate \eqref{eq:DM:Gamma} forces the mixing angle $\alpha_{13}$ to be $\sin \alpha_{13}\lesssim 10^{-11}$ to achieve the required large $v_R$ VEV. For the considered DM mass range, $10$~MeV and $1$~GeV, the decay rates \eqref{eq:DM:Gammaf} need an additional suppression which can be implemented by choosing $\mathbf{m}_D\lesssim 10$~GeV. In this regime, the largest scalar couplings of $R$  with other fields are typically of order $10^{-11}$, which also limits the DM interactions with the remaining particle content as this is set to the same magnitude by the small lepton number breaking conditions \eqref{eq:smalllep1}-\eqref{eq:smalllep4}. With all annihilation channels heavily suppressed, a freeze-out scenario inevitably produces an overabundance of DM.

In the freeze-in scenario \cite{Hall:2009bx}, however, the observed DM relic density can be achieved. Some benchmark points for which the correct DM abundance is obtained via freeze-in can be found in Table~\ref{tab:freezeout}. DM-Yukawa interactions $\mathbf{Y}_S=\boldsymbol{\mu}/v_R$ are suppressed not only by large values of $v_R$, but also by a small $\boldsymbol{\mu}$ scale that is assumed to be no larger than $10$~eV. The resulting DM-Yukawa interactions are then of the order of $10^{-22}$, hence they are negligible if compared to those proceeding in the scalar sector and the computation can thus be simplified by disregarding the  interactions with new fermions. In most of the considered configurations, $\lambda_{HJ}$ is the dominant portal between visible and dark sectors. As a consequence, we can simplify further our computations by fixing $\lambda_J = \lambda_R$ and $\lambda_{\rho J} = \lambda_{\rho R}$ and have the value of $\lambda_{RJ}$ to deviate from $2\lambda_{R}$ to obtain a desirable DM mass $m_J$. Then, $\lambda_{HJ}$ can be set so as to reproduce the observed DM relic density, as long as $\lambda_{HJ}$ remains between $0.82\lambda_{HR}<\lambda_{HJ}< 1.22\lambda_{HR}$ to satisfy the condition~\eqref{eq:smalllep1}. The DM mass we scan is limited to the range from $10$~MeV to $1$~GeV. Higher DM masses are less desirable due to the increase in the DM decay rate to light neutrinos~\eqref{eq:DM:Gamma} that they induce, yielding the strict constraint in~\cite{Arguelles:2022nbl}.

\subsection{Inflation} 
\label{sec:inflation}
Since the model contains several scalars, it also theoretically contains several candidates to drive the cosmological inflation. 
In the following, we focus our discussion on slow-roll single field inflation, leaving the exploration of alternative scenarios, such as constant-roll or multi-field inflation, to future works. We start with the Jordan frame action
\begin{equation}
S = \!\! \int \!\! d^4x \sqrt{-g^J}\left(-\frac{\MP^2}{16 \pi}f(s)\mathcal{R}^J + \frac{(\partial s)^2}{2}  - V(s) \right) ,
\label{eq:JframeL}
\end{equation}
where $\MP$ is the Planck mass, $\mathcal{R}^J$ is the Jordan frame Ricci scalar, $f(s)$ is the non-minimal coupling to gravity and $V(s)$ is the potential of the inflaton scalar $s$. 
It is possible to perform the inflationary computations in the Jordan frame, however, since cosmological perturbations are invariant under frame transformations (e.g \cite{Prokopec:2013zya,Jarv:2016sow}), it is more convenient to perform the analysis in the Einstein frame. This frame is reached via the Weyl transformation
\begin{eqnarray}
\label{eq:gE}
g^E_{\mu \nu} = f(s) \ g^J_{\mu \nu} \, , 
\end{eqnarray}
that yields the Einstein frame action
\begin{equation}
S = \!\! \int \!\! d^4x \sqrt{-g^E}\left(-\frac{\MP^2}{2} \mathcal{R}^E + \frac{(\partial \chi)^2}{2}  - U(\chi) \right) ,
\label{eq:L:classic}
\end{equation}
where the scalar potential $U(\chi)$ is given by
\begin{equation}
U(\chi) = \frac{V(s(\chi))}{f^{2}(s(\chi))} \, .
\label{eq:U:general}
\end{equation}
The canonically normalized field $\chi$ depends on the non-minimal function $f(s)$ and on the assumed gravity formulation.
In the usual metric formulation we have
\begin{equation}
\frac{\partial \chi}{\partial s} = \sqrt{\frac{3}{2}\left(\frac{\MP}{f}\frac{\partial f}{\partial s}\right)^2+\frac{1}{f}} \, ,  
  \label{eq:dphim}
\end{equation}
while for the Palatini formulation\footnote{More details about the Palatini formulation of non-minimal gravity can be found in e.g. \cite{Koivisto:2005yc,Bauer:2008zj,Gialamas:2023flv} and Refs. therein.} we have
\begin{equation}
\frac{\partial \chi}{\partial s} = \sqrt{\frac{1}{f}}  \, .  
  \label{eq:dphiP}
\end{equation}
Assuming slow-roll inflation, the evolution of the system can be described in the Einstein frame by the potential slow-roll parameters 
\begin{equation}
\epsilon_U = \frac{1}{2}M_{\rm P}^2 \left(\frac{1}{U}\frac{{\rm d}U}{{\rm d}\chi}\right)^2 \,, \quad
\eta_U = M_{\rm P}^2 \frac{1}{U}\frac{{\rm d}^2U}{{\rm d}\chi^2} \,.
\end{equation}
When $\epsilon_U \ll 1$, inflation takes place and the consequent expansion of the Universe can be measured in a number of $e$-folds
\begin{equation}
N_e = \frac{1}{M_{\rm P}^2} \int_{\chi_f}^{\chi_N} {\rm d}\chi \, U \left(\frac{{\rm d}U}{{\rm d} \chi}\right)^{-1},
\label{eq:Ne}
\end{equation}
where the field value at the end of inflation, $\chi_f$, is determined by $\epsilon_U(\chi_f) = 1$.  
The field value $\chi_N$ at the time a given scale left the horizon is given by the corresponding $N_e$. 
Other two important observables, i.e. the scalar spectral index and the tensor-to-scalar ratio are, respectively, written in terms of the slow-roll parameters as
\begin{eqnarray}
n_s &\simeq& 1+2\eta_U-6\epsilon_U, \label{eq:ns} \\
r &\simeq& 16\epsilon_U \, \label{eq:r} .
\end{eqnarray}
To reproduce the correct amplitude for the curvature power spectrum, the potential has to satisfy \cite{Planck2018:inflation}
\begin{equation}
\label{eq:As:constraint}
\ln \left(10^{10} A_s \right) = 3.044 \pm 0.014   \, ,
\end{equation}
where
\begin{equation}
 A_s = \frac{1}{24 \pi^2 \MP^4} \frac{U(\chi_N)}{\epsilon_U(\chi_N)} \label{eq:As} \, .
\end{equation}
Satisfying this last constraint usually
requires a rather flat inflaton potential. 

Therefore, the most natural inflaton candidate appears to be the scalar aligned with the flat direction of the scalar potential, i.e. the dilaton $\varphi$.  Unfortunately, the numerical studies of Section \ref{sec:dm} show that the quartic self-dilaton coupling is much smaller than $10^{-13}$ around the inflation scale,\footnote{A naive way to understand the issue is the following: as shown in Section \ref{sec:dm}, a correct dark matter relic density is only achieved when the scalar portal couplings are no bigger than $10^{-11}$. Therefore, we can naively estimate the quartic coupling of the dilaton to be of the order of $(10^{-11})^2 = 10^{-22}$. As mentioned in the main text, such a value is too small to satisfy the constraints coming from \cite{Planck2018:inflation}.} implying a scalar amplitude $A_s$ smaller than the measured value \cite{Planck2018:inflation}. The possible presence of a non-minimal coupling to curvature cannot solve this issue as it can only lead to even smaller scalar amplitudes. Similar issues hold for $J$, hence  the only remaining viable candidates are the Higgs boson or $\rho$. Since the former has been extensively studied in the context of inflation (e.g. \cite{Bezrukov:2007ep,Karananas:2022byw} and Refs. therein), we focus our efforts on $\rho$ identifying it, henceforth, with $s$.

\begin{figure}[t]
\begin{center}
  \includegraphics[width=0.49\textwidth]{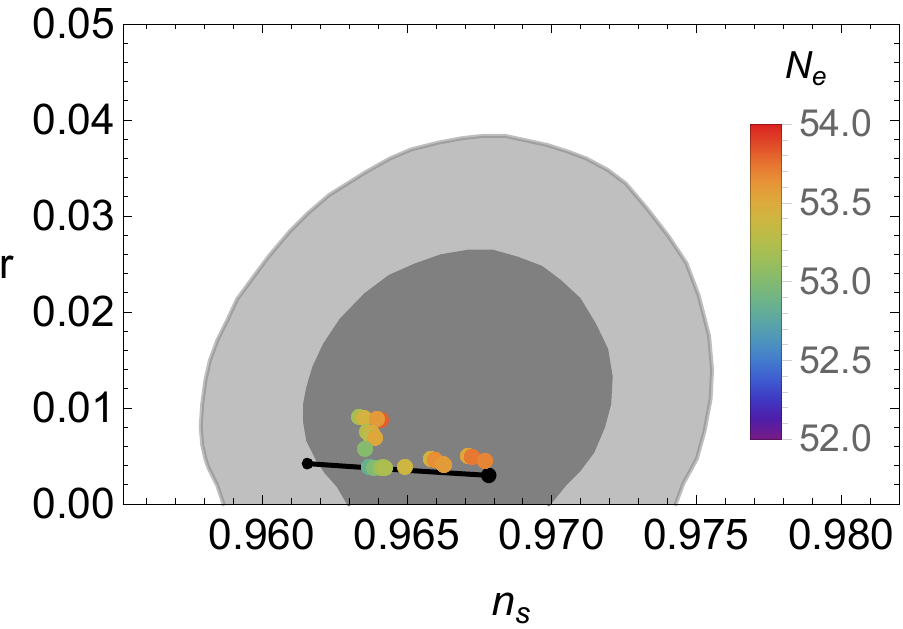}
  \includegraphics[width=0.49\textwidth]{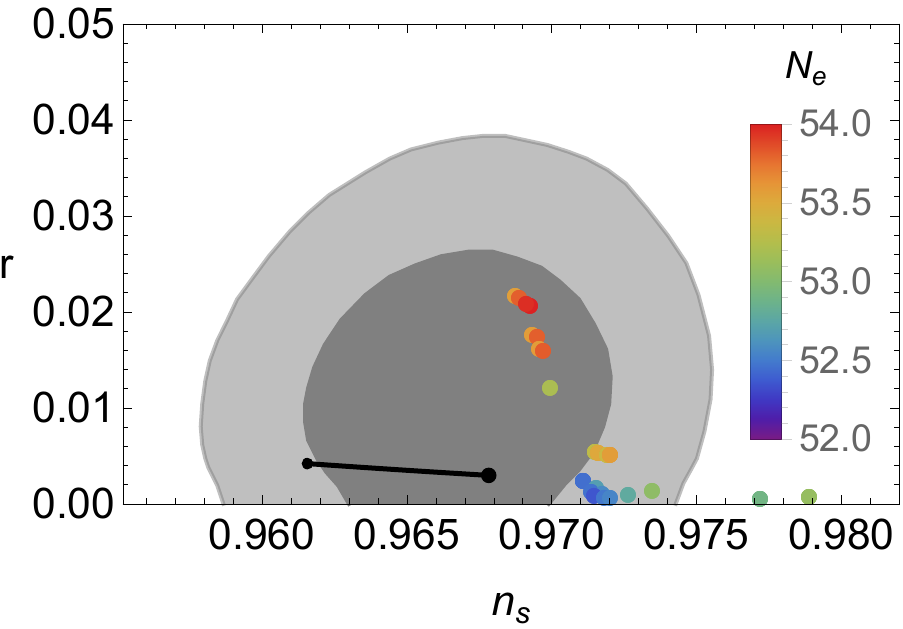}
\caption{$r$ vs. $n_s$ for the metric formulation of non-minimal gravity (left) and the Palatini formulation (right). The colour code is given in legend on the right side of the plot. In black are given the predictions for Starobinsky inflation for $N_e$ between 50 and 60. 
The gray areas represent the 1,2$\sigma$ allowed regions coming  from  the latest combination of Planck, BICEP/Keck and BAO data \cite{BICEP:2021xfz}. }
\label{fig:rvsns}
\end{center}
\end{figure}

As a proof of concept, we are going to consider scenarios in which the mixing between $\rho$ and the other scalar fields is minimal (as supported by current data), thereby avoiding all the problems arising from the diagonalization of the scalar sector.\footnote{We expect the inflationary predictions resulting from more general configurations to be  similar to the simpler case we discuss (see, e.g. \cite{Gialamas:2021enw}).} Therefore, the inflaton potential that we adopt in this article is simply the well-known Coleman-Weinberg (CW) type. This potential has been thoroughly  studied (e.g. \cite{Linde:1981mu, Albrecht:1982wi,Kannike:2014mia}) and known to be model-independently disfavoured by the latest constraints \cite{BICEP:2021xfz}. However, it can still be viable if we consider a non-minimal coupling of the inflaton to gravity of the Higgs-inflation type (e.g. \cite{Bezrukov:2007ep}): $f(s) = 1+ \xi s^2$. Note that a non-minimal coupling will, in any case, arise from RGE running. 
These frameworks, falling in the category of the non-minimal CW inflation models, usually yield inflationary predictions within the allowed ranges (see e.g. \cite{Marzola:2016xgb,Kannike:2018zwn,Racioppi:2018zoy,Racioppi:2019jsp}). As the corresponding formulas for the inflationary parameters are very cumbersome, they are relegated to the Appendix \ref{appendix:inflation}.
The results of our setup are given in Figure~\ref{fig:rvsns}, where we plot $r$ vs. $n_s$ for the metric formulation of non-minimal gravity (left) and the Palatini formulation (right). The gray areas represent the 1,2$\sigma$ allowed regions indicated by the latest combination of Planck, BICEP/Keck and BAO data \cite{BICEP:2021xfz}. In black are given the predictions for Starobinsky inflation for a number of $e$-folds, $N_e$, between 50 (small end) and 60 (larger end). The solutions of our model are displayed using a rainbow colour code which indicates the predicted number of $e$-folds, ranging from about 52 (violet) to 54 (red). 
$N_e$ has been computed by considering a reheating driven by the inflaton decay into SM particles proceeding via mixing with the Higgs boson, in the same fashion as \cite{Kannike:2018zwn}. 
 
In the Palatini formulation, all the points are quite well displaced from the Starobinsky limit. In the metric case, instead, only a subset of them is clearly distinguishable from the predictions of Starobinsky inflation. When the displacement takes place, for the metric case this is due to the effect of the radiative corrections, while for the Palatini one it is a combined effect of radiative corrections and of the chosen gravity formulation. We can see that the metric case is statistically more favoured than the Palatini one, as all the points fall within the 1$\sigma$ range indicated by the experiments. On the other hand, the Palatini completion sees most of its predictions fall outside of the 1$\sigma$ contour, with a relevant number of points that also exceed the 2$\sigma$ limit. The Palatini approach also provides a few points characterised by a substantial value of $r$, which fall within the reach of the dedicated next-generation experiments.

\section{Conclusions}
\label{sec:conclusions}

We have studied a classically scale-invariant realization of the inverse seesaw model where, in addition to the Higgs doublet $H$, the scalar sector contains a complex singlet $\sigma$ and a real singlet $\rho$. The VEVs of these scalars give rise to the neutrino mass matrix. The imaginary part of $\sigma$, identified with the Majoron $J$, acquires a small mass due to the explicit breaking of the lepton number induced by the scalar potential. 

We performed a Markov Chain Monte Carlo scan over the parameter space, taking into account experimental constraints affecting the Higgs boson signal strengths, its invisible decay width and neutrino masses, as well as theoretical bounds due to perturbativity and vacuum stability. We show that in a well-defined part of the parameter space satisfying these experimental constraints, the couplings remain perturbative up to the Planck scale and the scalar potential is bounded from below. In this way, we determine the parameter space that gives rise to the observed light neutrino mass pattern, as seen in figures~\ref{fig:c12s23} and \ref{fig:M1vsoth}.

Next, we re-analyse the selected parameter space to explain dark matter and inflation. We find that ensuring the stability of the Majoron on cosmological timescales, as necessary for a viable dark matter candidate, requires a sizeable dilaton VEV. Such a large VEV yields a suppression of DM couplings to the visible sector which precludes the possibility of matching the observed DM abundance via the freeze-out mechanism. The required relic density can still be produced via freeze-in through interactions in the scalar sector. The required mixing with the Higgs boson complies with current collider bounds.

We find that among the remaining scalar fields contained in the model, the real singlet $\rho$ and the Higgs boson can also play the role of the inflaton. In this article we concentrated our efforts on $\rho$ and studied the corresponding inflationary phenomenology under both the metric and Palatini formulations of non-minimal gravity. Our results show that the metric case is statistically favoured over the Palatini completion. However, both the cases yield a subset of testable configurations with a  the tensor-to-scalar ratio around (or larger than) $r \sim 0.01$, allowing for an experimental test in future experiments such as LiteBIRD \cite{LiteBIRD:2020khw}, and another subset with a predicted tensor-to-scalar ratio of the order of $r \sim 10^{-3}$, testable in principle by future planned satellite missions such as PICO \cite{NASAPICO:2019thw}. The remaining allowed results of the Palatini formulation, instead, unfortunately remain below the projected sensitivities of also these experiments. In the figures~\ref{fig:c12s23} and \ref{fig:M1vsoth}, the points that satisfy both dark matter and inflationary constraints are marked with a red star.

In conclusion, the proposed classically scale-invariant inverse seesaw model can explain the smallness of neutrino masses, provide a viable dark matter candidate and give origin to the a period of cosmological inflation compatible with current data.

\appendix

\section{Vacuum stability conditions}
\label{sec:vacuumstab}

We derive the vacuum stability or bounded-from-below conditions for the scalar potential \eqref{eq:V:rp} from copositivity~\cite{Kannike:2012pe}:
\begin{equation}
\begin{split}
  \lambda_{H} > 0, \quad \lambda_{\rho} > 0, \quad \lambda_{R} > 0, \quad \lambda_{J} &> 0,
  \\
  \bar{\lambda}_{H\rho} = \frac{1}{2} \lambda_{H\rho} + \sqrt{\lambda_{H} \lambda_{\rho}} > 0,
  \quad
  \bar{\lambda}_{HR} = \frac{1}{2} \lambda_{HR} + \sqrt{\lambda_{H} \lambda_{R}} &> 0,
  \\
  \bar{\lambda}_{HJ} = \frac{1}{2} \lambda_{HJ} + \sqrt{\lambda_{H} \lambda_{J}} > 0,
  \quad
  \bar{\lambda}_{\rho R} = \frac{1}{2} \lambda_{\rho R} + \sqrt{\lambda_{\rho} \lambda_{R}} &> 0,
  \\
  \bar{\lambda}_{\rho J} = \frac{1}{2} \lambda_{\rho J} + \sqrt{\lambda_{\rho} \lambda_{J}} > 0,
  \quad
  \bar{\lambda}_{RJ} = \frac{1}{2} \lambda_{RJ} + \sqrt{\lambda_{R} \lambda_{J}} &> 0,
  \\
  \sqrt{\lambda_{H} \lambda_{\rho} \lambda_{R}} + \lambda_{H\rho} \sqrt{\lambda_{R}}
  + \lambda_{HR} \sqrt{\lambda_{\rho}} + \lambda_{\rho R} \sqrt{\lambda_{H}}
  + \sqrt{2 \bar{\lambda}_{H\rho} \bar{\lambda}_{HR} \bar{\lambda}_{\rho R}} &> 0,
  \\
  \sqrt{\lambda_{H} \lambda_{\rho} \lambda_{J}} + \lambda_{H\rho} \sqrt{\lambda_{J}}
  + \lambda_{HJ} \sqrt{\lambda_{\rho}} + \lambda_{\rho J} \sqrt{\lambda_{H}}
  + \sqrt{2 \bar{\lambda}_{H\rho} \bar{\lambda}_{HJ} \bar{\lambda}_{\rho J}} &> 0,
  \\
  \sqrt{\lambda_{\rho} \lambda_{R} \lambda_{J}} + \lambda_{\rho R} \sqrt{\lambda_{J}}
  + \lambda_{\rho J} \sqrt{\lambda_{R}} + \lambda_{RJ} \sqrt{\lambda_{\rho}}
  + \sqrt{2 \bar{\lambda}_{\rho R} \bar{\lambda}_{\rho J} \bar{\lambda}_{RJ}} &> 0,
  \\
  \det (\mathbf{\Lambda}) > 0 \; \lor \; \text{some element(s) of } \adj (\mathbf{\Lambda}) &< 0,
\end{split}
\end{equation}
where the last condition, obtained from the Cottle-Habetler-Lemke theorem~\cite{Cottle1970295}, is not given in full. The adjugate matrix $\adj (\mathbf{A})$ of a matrix $\mathbf{A}$ is the transpose of the cofactor matrix of $\mathbf{A}$. It is defined through the relation $\mathbf{A} \adj (\mathbf{A}) = \det (\mathbf{A}) \mathbf{I}$.

\section{\texorpdfstring{$\beta$}{Beta}-functions}
\label{sec:betafunct}

We have calculated the $\beta$-functions with the {\tt PyR@TE 3} code~\cite{Sartore:2020gou}. The one-loop $\beta$-functions for the gauge, Yukawa (we neglect the down-type quark and lepton and up-type, save the top quark Yukawa coupling) and scalar quartic couplings are given by 
\begin{align}
  16 \pi^{2} \beta_{g'} &= \frac{41}{6} g^{\prime 3},
  \\
  16 \pi^{2} \beta_{g} &= -\frac{19}{6} g^{3},
  \\
  16 \pi^{2} \beta_{g_{3}} &= -7 g_{3}^{3},
  \\
  16 \pi^{2} \beta_{y_{t}} &= \frac{9}{2} y_{t}^{3} 
  + \tr\left(Y_D Y_D^{\dagger} \right) y_{t}
  - \frac{17}{12} g_1^{2} y_{t} - \frac{9}{4} g_2^{2} y_{t} - 8 g_3^{2} y_{t},
  \\
  16 \pi^{2} \beta_{Y_{D}} &= \frac{3}{2} Y_D Y_D^{\dagger} Y_D + \frac{1}{2} Y_D Y_{NS} Y_{NS}^{\dagger}
  + 3 y_{t}^2 Y_D + \tr\left(Y_D Y_D^{\dagger} \right) Y_D 
  \notag
  \\
  &- \frac{3}{4} g_1^{2} Y_D -\frac{9}{4} g_2^{2} Y_D,
  \\
  16 \pi^{2} \beta_{Y_{S}} &= \frac{1}{2} Y_{NS}^{T} Y_{NS}^{*} Y_{S} 
  + \frac{1}{2} Y_{S} Y_{NS}^{\dagger} Y_{NS} + 4 Y_{S} Y_{S}^{*} Y_{S}
  + 2 \tr\left(Y_{S} Y_{S}^{*} \right) Y_{S},
  \\
  16 \pi^{2} \beta_{Y_{NS}} &= Y_D^{\dagger} Y_D Y_{NS}+ 3 Y_{NS} Y_{NS}^{\dagger} Y_{NS}
  + 2 Y_{NS} Y_{S}^{*} Y_{S} + 2 \tr\left(Y_{NS} Y_{NS}^{\dagger} \right) Y_{NS},
  \\
  16 \pi^{2} \beta_{\lambda_{H}} &= \frac{3}{8} g^{\prime 4} \frac{3}{4} g^{\prime 2} g^{2}
  + \frac{9}{8} g^{2} + \lambda_{H} (24 \lambda_{H} - 3 g^{\prime 2} - 9 g^{2} + 12 y_{t}^{2} + 4 \tr Y_{D}^{\dagger} Y_{D})
  \notag
  \\
  &+ \frac{1}{2} \lambda_{HR}^{2} + \frac{1}{2} \lambda_{HJ}^{2} + \frac{1}{2} \lambda_{H\rho}^{2} - 2 \tr Y_{D}^{\dagger} Y_{D} Y_{D}^{\dagger} Y_{D} - 6 y_{t}^{4},
  \\
  16 \pi^{2} \beta_{\lambda_{R}} &= 18 \lambda_{R}^{2} + 2 \lambda_{HR}^{2} + \frac{1}{2} \lambda_{RJ}^{2} + \frac{1}{2} \lambda_{\rho R}^{2} + 8 \lambda_{R} \tr Y_{S}^{*} Y_{S} - 16 \tr Y_{S}^{*} Y_{S} Y_{S}^{*} Y_{S},
  \\
  16 \pi^{2} \beta_{\lambda_{J}} &=  18 \lambda_{J}^{2} + 2 \lambda_{HJ}^{2} + \frac{1}{2} \lambda_{RJ}^{2} + \frac{1}{2} \lambda_{\rho J}^{2} + 8 \lambda_{J} \tr Y_{S}^{*} Y_{S} - 16 \tr Y_{S}^{*} Y_{S} Y_{S}^{*} Y_{S},
  \\
  16 \pi^{2} \beta_{\lambda_{RJ}} &= \lambda_{RJ} (4 \lambda_{RJ} + 6 \lambda_{R} + 6 \lambda_{J} + 8 \tr Y_{S}^{*} Y_{S}) + 4 \lambda_{HR} \lambda_{HJ} + \lambda_{\rho R} \lambda_{\rho J} 
  \notag 
  \\
  &- 32 \tr Y_{S}^{*} Y_{S} Y_{S}^{*} Y_{S},
  \\
  16 \pi^{2} \beta_{\lambda_{\rho}} &= 18 \lambda_\rho^2 + 2 \lambda_{H\rho}^2 + \frac{1}{2} \lambda_{\rho R}^2
   + \frac{1}{2} \lambda_{\rho J}^2 + 8 \lambda_\rho \tr Y_{NS}^* Y_{NS} - 8 \tr Y_{NS}^* Y_{NS} Y_{NS}^* Y_{NS}, \label{eq:beta:lambda:rho}
  \\
  16 \pi^{2} \beta_{\lambda_{HR}} &= \lambda_{HR} \left(-\frac{3}{2} g^{\prime 2} - \frac{9}{2} g^{2} + 4 \lambda_{HR}  + 12 \lambda_{H} + 6 \lambda_{R} + 6 y_{t}^{2}  + 2 \tr Y_{D}^{\dagger} Y_{D} + 4 \tr Y_{S}^{*} Y_{S}  \right) 
  \notag
  \\
  &+ \lambda_{RJ} \lambda_{HJ} + \lambda_{H\rho} \lambda_{\rho R},
  \\
  16 \pi^{2} \beta_{\lambda_{HJ}} &= \lambda_{HJ} \left(-\frac{3}{2} g^{\prime 2} - \frac{9}{2} g^{2} + 4 \lambda_{HJ}  + 12 \lambda_{H} + 6 \lambda_{J} + 6 y_{t}^{2}  + 2 \tr Y_{D}^{\dagger} Y_{D} + 4 \tr Y_{S}^{*} Y_{S}  \right) 
  \notag
  \\
  &+ \lambda_{RJ} \lambda_{HR} + \lambda_{H\rho} \lambda_{\rho J},
  \\
  16 \pi^{2} \beta_{\lambda_{\rho R}} &= \lambda_{\rho R} (4 \lambda_{\rho R} + 6 \lambda_{R} + 6 \lambda_{\rho} + 4 \tr Y_{S}^{*} Y_{S} + 4 \tr Y_{NS}^{\dagger} Y_{NS})
  + 4 \lambda_{\rho} \lambda_{HR} + \lambda_{RJ} \lambda_{\rho J}
 \notag
 \\
  &- 32 \tr Y_{NS}^{\dagger} Y_{NS} Y_{S}^{*} Y_{S},
  \\
  16 \pi^{2} \beta_{\lambda_{\rho J}} &= \lambda_{\rho J} (4 \lambda_{\rho J} + 6 \lambda_{J} + 6 \lambda_{\rho} + 4 \tr Y_{S}^{*} Y_{S} + 4 \tr Y_{NS}^{\dagger} Y_{NS})
  + 4 \lambda_{\rho} \lambda_{HJ} + \lambda_{RJ} \lambda_{\rho R}
 \notag
 \\
  &- 32 \tr Y_{NS}^{\dagger} Y_{NS} Y_{S}^{*} Y_{S}.
\end{align}

\section{Flat direction}
\label{sec:flat:direction}

Any tree-level potential with biquadratic dependence on scalar fields can be written as
\begin{equation}
  V = \left(\mathbf{\Phi}^{\circ 2}\right)^T \mathbf{\Lambda} \mathbf{\Phi}^{\circ 2},
\end{equation} 
where the field vector $\mathbf{\Phi}$ collects all the real components of fields and the matrix $\mathbf{\Lambda}$ contains scalar quartic couplings.

The Hadamard product is defined as the element-wise product of two tensors of same dimensions, for instance: $(\mathbf{A} \circ \mathbf{B})_{ij} = A_{ij} B_{ij}$ for $\mathbf{A}$ and $\textbf{B}$ matrices. Similarly, the Hadamard power of a matrix is defined as $(\mathbf{A}^{\circ n})_{ij} = A^n_{ij}$ and so the Hadamard square of the field vector $\mathbf{\Phi}$ simply is $(\mathbf{\Phi}^{\circ 2})_{i} = (\mathbf{\Phi}_{i})^{2}$. The norm of $\mathbf{\Phi}$ is expressed via its Hadamard square as 
\begin{equation}
    \mathbf{\Phi}^T \mathbf{\Phi} = \mathbf{e}^T \mathbf{\Phi}^{\circ 2},
\end{equation}
where $\mathbf{e} = (1, \ldots, 1)$ is a vector of ones. Restricting the potential to the field vector $\mathbf{N}$ on the unit hypersphere with a Lagrange multiplier $\lambda$ gives
\begin{equation}
    V(\mathbf{N}, \lambda) = (\mathbf{N}^{\circ 2})^T \mathbf{\Lambda} \mathbf{N}^{\circ 2} 
    + \lambda (1 - \mathbf{e}^T \mathbf{N}^{\circ 2}).
\end{equation}
The minimum equations for $\mathbf{N}$ and $\lambda$ are
\begin{equation}
    2 \mathbf{N} \circ (2 \mathbf{\Lambda} \mathbf{N}^{\circ 2} - \lambda \mathbf{e}) = \mathbf{0},
    \qquad 
    \mathbf{e}^T \mathbf{N}^{\circ 2} = 1.
\end{equation}
Assuming that all elements of $\mathbf{N}$ are non-zero, we obtain
\begin{equation}
   2  \mathbf{\Lambda}\mathbf{N}^{\circ 2} = \lambda \mathbf{e}.
\label{eq:biq:phi:sq:min:lambda}
\end{equation}
Multiplying both sides of Eq.~\eqref{eq:biq:phi:sq:min:lambda} from the left by $(\mathbf{N}^{\circ 2})^{T}$ and using the unit length constraint yields
\begin{equation}
  \lambda = 2 (\mathbf{N}^{\circ 2})^T \mathbf{\Lambda} \mathbf{N}^{\circ 2} = 2 V(\mathbf{N}).
\end{equation}
Eq.~\eqref{eq:biq:phi:sq:min:lambda} then gives  
\begin{equation}
  \mathbf{\Lambda} \mathbf{N}^{\circ 2}
  = \left[(\mathbf{N}^{\circ 2})^T \mathbf{\Lambda} \mathbf{N}^{\circ 2} \right] \mathbf{e}
  \equiv V(\mathbf{N}) \, \mathbf{e},
\label{eq:biq:phi:sq:min:V}
\end{equation}
which we solve by the ansatz
\begin{equation}
  \mathbf{N}^{\circ 2} = C \adj(\mathbf{\Lambda}) \, \mathbf{e},
\label{eq:biq:ansatz:sol}
\end{equation}
where $C$ is a real normalization constant. Inserting Eq.~\eqref{eq:biq:ansatz:sol} into Eq.~\eqref{eq:biq:phi:sq:min:V}, we obtain
\begin{equation}
  C = \frac{1}{\mathbf{e}^T \! \adj(\mathbf{\Lambda}) \, \mathbf{e}},
 \label{eq:biq:ansatz:norm}
\end{equation}
which normalizes $\mathbf{N}$ to unity. 
Thus, the Hadamard square of the unit vector in the flat direction of the scalar potential is given by
\begin{equation}
  \mathbf{n}^{\circ 2} = \frac{\adj(\mathbf{\Lambda}) \, \mathbf{e}}{\mathbf{e}^T \! \adj(\mathbf{\Lambda}) \, \mathbf{e}},
\label{eq:flat:direction:sol:app}
\end{equation} 
where $\adj(\mathbf{\Lambda})$ is the adjugate of the coupling matrix, satisfying $\adj(\mathbf{\Lambda}) \mathbf{\Lambda} = \det (\mathbf{\Lambda}) \mathbf{I}$.

\section{Inflationary parameters} \label{appendix:inflation}
Here we present extended equations for the inflationary parameters. For the metric formulation we obtain
\begin{eqnarray}
    r &=& \frac{\left(M_{\rm P}^2 \left(4 \bar{\lambda }_{\rho }+\beta _{\lambda _{\rho }}\right)+4 M_{\rm P}^2 \beta _{\lambda _{\rho }} \ln
   \left(\frac{\rho_N }{M_{\rm P}}\right)+8 \pi  \xi  \rho_N ^2 \beta _{\lambda _{\rho }}\right)^2}{\pi  \rho_N ^2 \left(M_{\rm P}^2+8 \pi 
   \xi  (6 \xi +1) \rho_N ^2\right) \left(\bar{\lambda }_{\rho }+\beta _{\lambda _{\rho }} \ln \left(\frac{\rho_N
   }{M_{\rm P}}\right)\right)^2} \, ,
    \\
    n_s &=& 1+ \frac{1}{8 \pi  \left(\rho_N 
   M_{\rm P}^2+8 \pi  \xi  (6 \xi +1) \rho_N ^3\right)^2} \times \nonumber\\
   &&\times \Bigg[-\frac{2 M_{\rm P}^2 \beta _{\lambda _{\rho }} \left(M_{\rm P}^2+8 \pi  \xi  \rho_N ^2\right) \left(5 M_{\rm P}^2+8 \pi  \xi  (36 \xi +5)
   \rho_N ^2\right)}{\bar{\lambda }_{\rho }+\beta _{\lambda _{\rho }} \ln \left(\frac{\rho_N }{M_{\rm P}}\right)} + \nonumber\\
   &&\qquad -\frac{3 \beta
   _{\lambda _{\rho }}^2 \left(M_{\rm P}^2+8 \pi  \xi  \rho_N ^2\right)^2 \left(M_{\rm P}^2+8 \pi  \xi  (6 \xi +1) \rho_N
   ^2\right)}{\left(\bar{\lambda }_{\rho }+\beta _{\lambda _{\rho }} \ln \left(\frac{\rho_N }{M_{\rm P}}\right)\right)^2}+ \nonumber\\
   && \qquad +8 M_{\rm P}^2
   \left(-8 \pi  \xi  (24 \xi +5) \rho_N ^2 M_{\rm P}^2-3 M_{\rm P}^4-128 \pi ^2 \xi ^2 (6 \xi +1) \rho_N ^4\right) \Bigg] \, ,
    \\
    A_s &=& \frac{32 \pi  \rho_N ^6 \left(M_{\rm P}^2+8 \pi  \xi  (6 \xi +1) \rho_N ^2\right) \left(\bar{\lambda }_{\rho }+\beta _{\lambda _{\rho }}
   \ln \left(\frac{\rho_N }{M_{\rm P}}\right)\right)^3}{3 \left(M_{\rm P}^2+8 \pi  \xi  \rho_N ^2\right)^2 \left(M_{\rm P}^2 \left(4
   \bar{\lambda }_{\rho }+\beta _{\lambda _{\rho }}\right)+4 M_{\rm P}^2 \beta _{\lambda _{\rho }} \ln \left(\frac{\rho_N
   }{M_{\rm P}}\right)+8 \pi  \xi  \rho_N ^2 \beta _{\lambda _{\rho }}\right)^2} \, ,
\end{eqnarray}
while for the Palatini case we have
\begin{eqnarray}
    r &=& \frac{\left(M_{\rm P}^3 \left(4 \bar{\lambda }_{\rho }+\beta _{\lambda _{\rho }}\right)+4 M_{\rm P}^3 \beta _{\lambda _{\rho }} \ln
   \left(\frac{\rho_N }{M_{\rm P}}\right)+8 \pi  \xi  \rho_N ^2 M_{\rm P} \beta _{\lambda _{\rho }}\right)^2}{\pi  \rho_N ^2 \left(M_{\rm P}^2+8 \pi
    \xi  \rho_N ^2\right)^2 \left(\bar{\lambda }_{\rho }+\beta _{\lambda _{\rho }} \ln \left(\frac{\rho_N
   }{M_{\rm P}}\right)\right)^2} \, ,
    \\
    n_s &=& 1-\frac{\frac{3 M_{\rm P}^2 \beta _{\lambda _{\rho }}^2}{\left(\bar{\lambda }_{\rho }+\beta _{\lambda _{\rho }} \ln
   \left(\frac{\rho_N }{M_{\rm P}}\right)\right)^2}+\frac{2 \beta _{\lambda _{\rho }} \left(8 \pi  \xi  \rho_N ^2 M_{\rm P}^2+5
   M_{\rm P}^4\right)}{\left(M_{\rm P}^2+8 \pi  \xi  \rho_N ^2\right) \left(\bar{\lambda }_{\rho }+\beta _{\lambda _{\rho }} \ln
   \left(\frac{\rho_N }{M_{\rm P}}\right)\right)}+\frac{24 M_{\rm P}^4}{M_{\rm P}^2+8 \pi  \xi  \rho_N ^2}}{8 \pi  \rho_N ^2} \, ,
    \\
    A_s &=& \frac{32 \pi  \rho_N ^6 \left(\bar{\lambda }_{\rho }+\beta _{\lambda _{\rho }} \ln \left(\frac{\rho_N }{M_{\rm P}}\right)\right)^3}{3
   \left(M_{\rm P}^3 \left(4 \bar{\lambda }_{\rho }+\beta _{\lambda _{\rho }}\right)+4 M_{\rm P}^3 \beta _{\lambda _{\rho }} \ln
   \left(\frac{\rho_N }{M_{\rm P}}\right)+8 \pi  \xi  \rho_N ^2 M_{\rm P} \beta _{\lambda _{\rho }}\right)^2} \, ,
\end{eqnarray}
where
\begin{eqnarray}
 \bar\lambda_\rho &=&\lambda _{\rho }+
  \frac{\lambda _{H \rho}^2 \ln \left(\frac{\lambda _{H \rho}}{2}\right)}{16 \pi ^2}+
 \frac{9   \lambda _{\rho }^2 \ln \left(3 \lambda _{\rho }\right)}{16 \pi ^2}+\frac{\lambda _{\rho J}^2 \ln
   \left(\frac{\lambda _{\rho J}}{2}\right)}{64 \pi ^2}+\frac{\lambda _{\rho R}^2 \ln \left(\frac{\lambda_{\rho R}}{2}\right)}{64 \pi ^2} \nonumber \\
  && -\frac{\tr Y_{NS}^* Y_{NS} Y_{NS}^* Y_{NS} \ln \left( \tr Y_{NS}^* Y_{NS} \right) }{4 \pi ^2} \, ,
\end{eqnarray}
and $\beta_{\lambda_\rho}$ is given in eq. \eqref{eq:beta:lambda:rho} and all the couplings are computed at the Planck scale $\MP$. The field value $\rho_N$ is computed from the number of $e$-folds $N_e$ in eq. \eqref{eq:Ne}. $N_e$ has been computed by considering a reheating driven by the inflaton decay into SM particles proceeding via mixing with the Higgs boson, in the same fashion as \cite{Kannike:2018zwn}. 

\acknowledgments

We would like to thank Lorenzo Calibbi for useful discussions. This work was supported by the Estonian Research Council grants MOBTT5, PRG356, PRG434 and PRG1055.

\bibliographystyle{JHEP}
\bibliography{CSIinvSS}

\end{document}